\documentstyle[aps,prd,epsfig]{revtex}

\newcommand{\be}{\begin{equation}}
\newcommand{\ee}{\end{equation}}
\newcommand{\bea}{\begin{eqnarray}}
\newcommand{\eea}{\end{eqnarray}}
\newcommand{\kp}{{\bf k}_\perp}
\newcommand{\bp}{{\bf p}}
\newcommand{\bk}{{\bf k}}
\newcommand{\bgam}{\mbox{\boldmath$\gamma$}}
\newcommand{\blam}{\mbox{\boldmath$\lambda$}}

\def\la{\mathrel{\mathpalette\fun <}}
\def\ga{\mathrel{\mathpalette\fun >}}
\def\fun#1#2{\lower3.6pt\vbox{\baselineskip0pt\lineskip.9pt
\ialign{$\mathsurround=0pt#1\hfil##\hfil$\crcr#2\crcr\sim\crcr}}}

\begin{document}

\title{Quark combinatorics and meson production ratios
in hadronic $Z^0$ decays}
\author{V. V. Anisovich, V. A. Nikonov}
\address{Petersburg Nuclear Physics Institute, Gatchina,
188300 Russia}
\author{and J. Nyiri}
\address{Research Institute for Particle and Nuclear Physics, Budapest,
Hungary}
\date{August 14, 2000}
\maketitle

\begin{abstract}
Rich experimental data accumulated in the past few years  in the hadronic
$Z^0$ decays allow one to check the quark combinatorics relations for
a new type of processes, namely: quark jets in the decays $Z^0 \to
q\bar q \to hadrons$.  In this paper we review quark
combinatorics rules for the yields of vector and pseudoscalar mesons,
$V/P$, in the central and fragmentation regions of the hadronic $Z^0$
decays.  It is stressed that in the central region a direct
verification of quark combinatorics rules is rather problematic because
of a considerable background related to the decay of highly excited
resonances; however such a verification is possible in the
fragmentation region, at $x_{hadron} \sim 0.5-1$, where the
contribution of resonance decays is suppressed due to the rapid
decrease of spectra with increasing $x_{hadron}$. It is shown
that the experimental data in the fragmentation region for
$\rho^0/\pi^0$ and $p/\pi^+$ are in a reasonable agreement with
the predictions of quark combinatorics.
 The ratios of the heavy meson yields,
$B^*/B$ and $D^*/D$, are also discussed: the data demonstrate a
good agreement with the quark combinatorial results.  We analyse the
structure of the suppression parameters for strange and heavy quark
productions in soft processes and estimate their order of magnitudes
in the multiperipheral processes; the ratio
$K^{\pm}/\pi^{\pm}$ at $x_{hadron}\sim 0.5-0.8$ and
production probabilities of the $c\bar c$
and $b \bar b$ mesons are in a good agreement with the estimates.

\end{abstract}

\section{Introduction}

The relations of quark combinatorics for hadron yields were originally based
on the qualitative treatment of a multiparticle production process
as a process of production of a "cloud" of non-correlated quarks and
antiquarks followed by their occasional fusion into mesons and baryons
\cite{1,2}. In this picture the quark combinatorics approach echoes
the statistical treatment of reactions of nucleus decays
\cite{2*}, and actually it was initiated by the success of the
statistical methods in the description of multinucleon processes.
There is one more common point in  treating multinucleon and
multiquark processes: the stage of final state interactions.
In multinucleon reactions the Watson--Migdal factor \cite{3*} leads
to an increase of the low-energy contributions in the nucleon--nucleon
final state interactions.
In multiquark reactions
the analogous effect of the $q\bar q$ and $qqq$ final state
interactions, as one can guess, should enlarge the production of the
comparatively low-mass mesons and baryons: in \cite{1*} a hypothesis
was made about the dominant contribution of several low-mass $q\bar
q$ and $qqq$ multiplets to the spectra of produced hadrons.

This particular story reflects a general trend of
how methods and approaches used in nuclear physics influenced the
physics of strong interactions. But, as time goes on, our understanding
of QCD, the theory of strong interactions, becomes deeper, and
now a new level of understanding of quark-gluon physics in the soft
interaction region forces us to review the relations of quark
combinatorics.

{\bf a) Quark combinatorics for hadron production.}

The rules of quark combinatorics,
as it was stressed above, were initially suggested
for finding probabilities of hadron yields  in the multiparticle
production processes \cite{1,2}.
Later, they were extended to
meson decay processes, see \cite{3}. In \cite{1,2} (see also
\cite{1*}) the
hadronic production processes were treated on the basis of the
following qualitative picture.
In the multiparticle production process
there is a stage of a cloud of constituent quarks. When joining each
other, they create mesons ($q\bar q$), baryons ($qqq$) and antibaryons
($\bar q\bar q\bar q$). The probabilities to form hadrons of
different sorts are defined by combinatorial rules which have been
obtained under the assumption that quarks and antiquarks of the sea are
produced uncorrelated and there is neither spin nor isospin alignment.

Our knowledge of the multiparticle production mechanism, i.e. the
structure of amplitudes responsible for the inclusive production,
is now, however,
more complete, thus allowing us to
investigate the basis of quark combinatorics at a new level.
In the present paper, we investigate\\
(i) the ratio of vector/pseudoscalar meson yields both for the central
and fragmentation regions of the quark jets in the hadronic $Z^0$ decays
(Sections 2 and 3, respectively), and\\
(ii) the structure of the suppression parameters for
production of strange and heavy quarks in the multiperipheral
processes (Section 4).

Time is ripe for this review,
for in the meantime quark combinatorics has become
widely used for the consideration of meson
resonances with masses 1.0-2.5 GeV, with a goal to determine the
quark-gluon content of these states by investigating the transitions
$resonance \to mesons$ \cite{4,5,6}. In the investigation of
exclusive decay processes \cite{4,5,6}, the qualitative picture of the
cloud structure of sea quarks is not used (that
makes the conclusions  more model-independent). However, the notion
"suppression parameter" for the production of
strange quarks plays  an important role ,
and it is the same as what is
used in multiparticle production processes.
Thus, we have to re-think
quark combinatorics on a new basis both
theoretically and experimentally.

{\bf b) Hadronic $Z^0$ decays.}

We discuss here the yields of vector $(\rho^0, \omega, \bar K^*)$ and
pseudoscalar $(\pi, K)$ mesons in the hadronic $Z^0$ decay. Currently
there exists rich experimental information on these processes, which are
determined by transitions $Z^0 \to q\bar q \to hadrons$. First, we
consider the production of light--flavour mesons, which
are created in the quark jets $Z^0 \to u\bar u$, $Z^0 \to d\bar d$, $Z^0
\to s\bar s$ with probabilities  \cite{PDG}:
\begin{equation}
u\bar u:d\bar d:s\bar s \simeq 0.26:037:0.37
\end{equation}
The large mass of the $Z^0$ boson makes it possible to observe in hadronic
decays $Z^0 \to q\bar q \to hadrons$ the characteristic features of
both multiparticle production (central region of quark jets) and meson
decay processes (fragmentation production).

The data accumulated by the Collaborations ALEPH \cite{ALEPH}, L3
\cite{10}, DELPHI \cite{11} and OPAL \cite{12} provide spectra of
vector and pseudoscalar mesons, $d\sigma/dx(Z^0 \to V+X)$ and
$d\sigma/dx(Z^0 \to P+X)$ in a broad interval of $x$, thus giving an
opportunity to compare them with the predictions of quark combinatorics.

{\bf c) Hadron production in the central region, and the
vector-pseudoscalar ratio. }

In Section 2 the hadron production in the central region in the quark jet is
considered.  We discuss the prompt vector meson $(V)$ and
pseudoscalar meson $(P)$ yields . For particles belonging to the same
$q\bar q$ multiplet we obtain:
\be \frac{V_{prompt}}{P_{prompt}}=3.
\ee
This is a well-known result of the quark combinatorics \cite{1,2,1*} for
hadron--hadron collisions. Hence, our analysis shows that the
ratio $V_{prompt}/P_{prompt}$ is the same
for pomeron ladder in hadron--hadron collisions and
for quark jets, despite of the different structure of the colour exchanges
in these processes.

As we have already said, the rules of quark combinatorics for
hadron-hadron collisions were introduced in \cite{1,2,1*}.
Investigations
of the QCD-pomeron \cite{13,14}
shed light on
 quark--gluon structure
of the multiperipheral ladder in hadron collisions and allowed us to
deal with meson yields in the central region on a new level.
Therefore,
in Appendix A we give a new analysis of the results of quark combinatorics
for $V_{prompt}/P_{prompt}$ at $x\sim 0$, using Lipatov's
pomeron \cite{14} as a guide.

{\bf d) Suppression of strange quark production.}

An important feature in hadron production is the suppression of the
production probability of the strange quark in the multiperipheral
ladder:
\be u\bar u:d\bar d:s\bar s=1:1:\lambda \ ,
\ee
with $0\le \lambda
\le 1$.  This topic is discussed at Section 4: it is shown that
$\lambda \simeq m^2/m^2_s$, where $m$ and $m_s$ are masses of
non-strange and strange constituent quarks, respectively.

This suppression results in a suppression of the production of
mesons containing strange quarks that is observed
in the hadronic $Z^0$ decays.

{\bf e) Soft colour neutralisation of quarks in the hadronic $Z^0$
decay.}

When considering central production at the decay $Z^0 \to q\bar q \to
hadrons$ we start with the standard mechanism of soft colour
neutralisation of
the outgoing quarks: newly-born quark--antiquark pairs are
produced in multiperipheral ladder (see Fig. 1a), which provides the
transfer of the colour from antiquark to quark. The discontinuity of
the self-energy diagram of Fig. 1b (determined by cutting through
hadronic states, dashed line in Fig. 1b) defines
the transition cross section $Z^0 \to
hadrons$, while the quark-gluon block inside the big quark loop
determines the confinement forces.

Likewise, the inclusive production cross section of the meson in the
central region is
provided by the discontinuity of the diagram of
Fig. 1c. The quark loop $meson\to q\bar q\to meson$
shown in Fig. 1c with the production of vector
or pseudovector mesons determines relative probabilities of these
particles.  The chain of the quark loops shown on Figs. 1b, 1c
and 1d (below we denote this chain as $A$) contains both colour singlet
$(c=1)$ and colour octet $(c=8)$ components: $A=A_1+A_8$. According to
the rules of $1/N$-expansion \cite{15}, the main contribution is due to
the octet component.
The idea that the quark leaves the confinement trap by
 a production of new
quark--antiquark pairs is rather old; it has been discussed long
ago (for example, see \cite{1*}, Sections 7 and 9, as well as
\cite{16}).  A new impulse in understanding the confinement mechanism
has been brought by Gribov \cite{17}. Following the ideas of \cite{17},
we use the jet structure shown in Fig. 1a assuming that
the $t$-channel exchange by quark is a constructive
element of the jet.

Spectroscopic calculations (see, for example, \cite{18}) support the
hypothesis about the scalar type of confinement forces; accordingly, we
assume that the chain $A$ realises the $t$-channel exchange
with $J^P=0^+$.

The calculation of the block for central meson
production  performed in Section 2 proves that equation (2) is
fulfilled, if the wave functions of mesons ($V$ and $P$) belonging to
the same multiplet are equal.

{\bf f)  Prompt hadron production and production due to the
decay of high resonances.}

The equality (2)  is valid for prompt meson production, while the
decays of highly excited states violate this ratio as is seen in the
experiment, providing the increase of the rate of
 light mesons due to resonance decay. As to $\rho/\pi$ and $K^*/K$,
the decays increase the contribution of pseudoscalar component.
According to \cite{ALEPH}, $\rho^0/\pi^0=0.15\pm 0.03$, and
$K^*/K=0.40\pm 0.06$, that indicates on a large contribution into
spectra from the decays of highly excited states.

The problem of the production and
decay of highly excited states in hadron--hadron
collisions had been discussed in \cite{19}. The conclusion was similar:
average multiplicities of the produced light mesons and baryons result
mainly from the cascade decays of the highly excited resonances.

The multiperipheral structure of the ladder allows us to estimate the
resonance masses which are important for meson production; they are
of the order or less than 2 GeV (Section 5)

The existence of the  decay of highly excited
resonances is a reality that one should take into account in the
verification of quark combinatorial rules. We discuss several ways
of solving this problem.

{\bf g) Heavy meson production in
$Z^0\to b\bar b$ and $Z^0\to c\bar c$ decays.}

One way is to check quark combinatorics for heavy
particle yields, where the  cascade multiplication is suppressed.
An ideal example could
be the production of mesons containing a $b$-quark.
In fact, for the beauty mesons the ratio
$B^*/B$ observed in the experiment agrees with (2). When the
lowest $S$-wave multiplet dominates in the production of heavy mesons,
then one has (provided the equation (2) is fulfilled): $B\simeq
B_{prompt}+B^*_{prompt}=4B_{prompt}$, and the ratio of vector
and pseudoscalar mesons is $B^*/B\simeq 0.75$.

At the experiment on $Z^0$ decay it was observed:  $B^*/B=0.771\pm
0.075$ \cite{20}, $0.72\pm 0.06$ \cite{21}, $0.76\pm 0.10$ \cite{22},
$0.76\pm 0.09$ \cite{23}; the mean value is $0.75\pm 0.04$.

For charmed mesons $D^*/D=0.60\pm 0.05$ \cite{24},
$0.62\pm 0.03$ \cite{25}, $0.57\pm 0.05$ \cite{26}. The mean value is
$0.61\pm 0.03$,which means a rise of the contribution from the
decay of the non-$S$-wave multiplets.

One should stress that the ratios $V/P$ for beauty  and
charmed mesons are saturated in
the fragmentation region due to the transitions $Z^0\to b\bar b$
and $Z^0\to c\bar c$. Therefore, in Section 3
we re-analyse quark combinatorics for the fragmentation region of the
hadronic $Z^0$ decay.

The production cross section of mesons in the fragmentation region is
determined by the discontinuity of the diagram of Fig. 1c (the cutting
of ladder diagram is shown by dashed line). Direct
calculations demonstate that (2) is fulfilled with rather good
accuracy for the fragmentation region as well, provided the wave
functions of vector and pseudoscalar mesons are equal.

{\bf h) Production of light-flavour mesons in the fragmentation region
of the hadronic $Z^0$ decay.}

Investigations of meson production in the fragmentation region open the
way to test the rules of quark combinatorics for light--flavour
hadrons, and to verify (2) in particular. As is said above,
in the spectra of light--flavour hadrons the contribution of the
component related to the decay of highly excited states dominates.
Still, in the case of jet processes this component dominates in the
central region, at $x\sim 0$, but
not in the fragmentation one. The hadronic spectra for jets are
maximal at $x\sim 0$, and they decrease rapidly with the growth of $x$.
As a result, the component which comes from a resonance decay decreases
quickly, because the decay products share
the value $x_{resonance}$,
thus entering the region of smaller $x$. In due course this effects
a fast growth of relative contributions from prompt particle
production. Therefore, the measurements of particle yields at $x\sim
0.5-1.0$ provide an opportunity for model--independent testing of
quark combinatorics.

In Section 3 we compare the ratios $\rho^0/\pi^0$ and $K^*/K$
measured in \cite{ALEPH} at large $x$ with quark combinatorics
predictions;
in Section 5 the baryon-meson ratio,  $p/\pi^+$, is discussed.
comparison demonstates reasonable agreement of the
experimental data with the predictions of quark combinatorics.

{\bf i) Heavy quark production suppression.}

Multiperipheral dynamics of the quark-gluon ladder in the
processes of Fig. 1 results in a strong suppression of production of
new heavy quark pairs, $Q\bar Q$. The suppression parameter
(relative probability) is of the order of $\lambda_Q  \simeq
m^2/(m_Q^2\ln^2(\Lambda^2/m_Q^2))$, where $m_Q$ is the mass of  heavy
quark and $\Lambda $ is the QCD scale parameter, $\Lambda \sim 200$
MeV. One has $\lambda_c \simeq 2.8 \cdot 10^{-3}$  and
 $\lambda_b \simeq 1.1\cdot 10^{-4}$.

The inclusive production of the $c\bar c$ and $b\bar b$ mesons in $Z^0$
decays agrees with this estimation (Section 4).

{\bf j) The ratio $\omega/\rho^0$ and interference of flavour
amplitudes at $x\sim 1$.}

The rules of quark combinatorics give us
$\omega_{prompt}/\rho^0_{prompt}=1$.
It can be proven that the decays of highly excited states effect weakly
this equality. Indeed, the masses of $\rho$ and $\omega$ are almost equal,
so the phase spaces of the decay processes influence equally the
probabilities of their yields. The experiment confirms this statement:
at $x\sim 0.1-0.2$ the value $\omega/\rho^0=1$ is observed. However, at
larger $x$ the experimental value is $\omega/\rho^0<1$. This reveals
the interference of flavour amplitudes. The fact is that the decay
vertices for $Z^0\to u\bar u$ and $Z^0\to d\bar d$ have opposite signs,
thus enhancing the coherent production of the $\rho^0$ meson and
suppressing the production of $\omega$ (Section 6).

 \section{Prompt production in the central region of the
quark jet:  the ratio $V/P=3$}

In this Section we discuss the prompt production of vector and
pseudoscalar mesons (for definiteness, $\rho$ and $\pi$) in the
central region of a quark jet. The production cross section is
determined by the discontinuity of the diagram shown in Fig. 1c;
it is re-drawn in Fig. 2a. A particular feature of
the production of $\rho$ and $\pi$ is the presence of a loop
diagram, which is shown separately by Fig. 2b. Below we
calculate these loop diagrams for $\rho$ and $\pi$ using
the spectral integration
technique which is discussed in detail in \cite{AMN,AMMP}. Within this
technique the loop diagrams are expressed in terms of the $\rho$ and
$\pi$ light--cone wave functions.

But first, let us present the result of our calculations.

Direct calculations lead to the following formulae for the inclusive
cross section of the  $\rho$ and $\pi$ mesons at $x\sim 0$:
\begin{eqnarray}
&& \frac{d\sigma}{dx}(Z^0\to\rho +X)=\frac1{16\pi^3}\int\limits^1_0
\frac{d\xi}{\xi(1-\xi)}\int d^2{\bf k}_\perp \psi^2_\rho(\xi,{\bf
k}_\perp)\cdot 3\Pi_Z(W_1^2,W_2^2) \nonumber\\
&& \frac{d\sigma}{dx}(Z^0\to\pi +X)=\frac1{16\pi^3}\int\limits^1_0
\frac{d\xi}{\xi(1-\xi)}\int d^2{\bf k}_\perp \psi^2_\pi(\xi,{\bf
k}_\perp)\cdot \Pi_Z(W_1^2,W_2^2) \ .
\label{2-1}
\end{eqnarray}
Here $\psi_{\rho}$ and $\psi_{\pi}$ are quark wave functions of
$\rho$ and $\pi$ mesons, $\xi$ and $\kp$ are
quark light--cone variables (the
fraction of momentum carried by a quark along the
$z$-axis and its momentum in the $(x,y)$-plane, respectively).
In (4) one can see explicit expressions related to the production of
$\rho$ and $\pi$ mesons. The rest
(contribution from the large quark loop as well as from ladder
diagrams) is designated in (4) as $\Pi_Z(W_1^2,W_2^2) $, which depends
on the invariant energies squared for the quark chains,
$W_1^2$ and $W_2^2$.
Multiperipheral kinematics give:
\be
W_1^2 \; W_2^2 \simeq \xi(1-\xi)(m^2+k^2_{\perp})M^2_Z \; .
\label{2-2}
\ee
Here $M_Z$ is the mass of the $Z^0$ boson.

The factor 3
in the $\rho$ production cross section is the result of summing over
polarizations of the vector particle.

Equation (4) directly demonstrates that, if quark wave functions of
$\rho$ and $\pi$ are identical, that is assumed by the quark multiplet
classification of these mesons, then at $x\sim 0$ we have $
d\sigma(Z^0\to\rho +X)/dx:d\sigma(Z^0\to\pi +X)/dx\ =\ 3:1\ $.
Let us stress again: the equation (4) as well as diagrams of Figs. 1c,
2a, 2b stand for promptly produced mesons.

\subsection{Spectral representation for the loop diagram of Fig. 2a }

The calculation of the diagram of Fig. 2a in terms of spectral
integration is performed in the following steps (see also
\cite{AMN,AMMP}): \\ (i)  quark loops in Fig. 2a are taken
energy-off-shell; \\
(ii) for the energy-off-shell quark loops
the discontinuities are calculated (corresponding
cuts are shown by the dotted lines I, II, III and IV); \\
(iii) the spectral integrals are determined by the discontinuities
being the integrands.

First, consider the double spectral integral which
corresponds to the cuts III and IV, which are  the
spectral integrals over effective masses squared,  $M^2$ and $M'^2$,
in the transitions $Z^0 \to q\bar q$ and
$q\bar q \to Z^0$:
$$
\int \limits_{4m^2}^{\infty}\frac {dM^2dM'^2}{\pi^2} \,
\frac{g_Z(M^2)g_Z(M'^2)}{(M^2-M_Z^2)(M'^2-M_Z^2)}\,
d\Phi(P_Z;p_1,p_2) d\Phi(P'_Z;p_3,p_4)\;
$$
\be
\times \; S_{Z}\; T(p;q_1,q_2) A(W^2_1,q^2_1)A(W^2_2,q^2_2) \; .
\label{2a-1}
\ee
Here $g_Z(M^2)$ is the
vertex function for the $Z^0\to q\bar q$ transition and $M_Z$ is the
$Z^0$ boson mass. The factors $d\Phi(P_Z;p_1,p_2)$
and $ d\Phi(P'_Z;p_3,p_4)$ are  phase spaces related to the
cuts III and IV. $S_{Z}$ is the spin factor for the big quark loop
in Fig.  2a. The amplitudes $A(W^2_i,q^2_i)$, ($i=1,2$), refer to the
quark-gluon chains with the $t$-channel scalar quantum numbers (see
discussion in Section 1.e), and $T(p;q_1,q_2)$ is the block which
corresponds to the small quark loop (the $q\bar q$ loop for production
of $\rho$ and $\pi$ mesons).

The characteristic feature of the spectral integral (6) is the large value
of $M_Z$. Because of that one can replace, with rather good accuracy,
the poles of the spectral integrand by half-residues:
\be
\frac1{M^2-M_Z^2} \to -i\pi \delta (M^2-M_Z^2), \qquad
\frac1{M'^2-M_Z^2} \to i\pi \delta (M'^2-M_Z^2) \; .
\label{2a-2}
\ee
Equation (6) stands for the discontinuity of the amplitude that
results in different signs for the half-residues in (7). Equation
(7) means that the block inside of the big quark loop can be
considered, with a good accuracy, as a block of the real process. This
is a well-known feature of the high-energy jets; below we use it
for estimating meson production amplitudes.

We are not especially interested in the spectral integral (6) determined
by the big quark loop, since it is quite common for the $\pi$ and
$\rho$ production. Our target is the spectral integral which
corresponds to the amplitude $T(p;q_1,q_2)$ that is the $\rho$ or $\pi$
meson production block. This block is shown separately in Fig. 2b.

\subsection{Calculation of the loop diagram of Fig. 2b }

The amplitude of the loop diagram of Fig. 2b represented as
a double dispersion integral is:
\be
{\mbox M} \ =\ T(p;q_1,q_2)2p_0(2\pi)^3\delta^{(3)}({\bf p}+{\bf
q}_1-{\bf q}_2 -{\bf p}')\ ,
\label{2b-1}
\ee
$$
T(p;q_1,q_2)\ =\int\limits^\infty_{4m^2}\frac{dsds'}{\pi^2}\int
d\phi \; \frac{G_{\rm meson}(s)}{s-\mu^2} \frac{G_{\rm
meson}(s')}{s'-\mu^2}\; S_{\rm meson}\ .
$$
Here $s$ and $s'$ are the invariant masses squared in the intermediate
$q\bar q$-states; $G_{\rm meson}$ is the vertex for the
$meson \to q\bar q$ transition. The ratio $G_{\rm meson}(s)/(s-\mu^2)$
determines the wave function of the produced meson
up to the spin factor,
and $S_{\rm meson}$ is the spin factor of the loop diagram 2b.

Let us explain the structure of the spectral integral (\ref{2b-1})
in  more detail.

The cut quark loop should be integrated over the phase space of the
intermediate $q\bar q$-states, $\Phi(P;k_1,k_2)$ and
$\Phi(P';k'_1,k'_2)$, with the invariant phase space determined as
\be
d\Phi(P;k_1,k_2)=\frac12\,\frac{d^3k_1}{(2\pi)^32k_{10}}\
\frac{d^3k_2}{(2\pi)^32k_{20}}\ (2\pi)^4\delta^{(4)}(P-k_1-k_2)\ .
\label{2b-2}
\ee
The integration is carried out taking into account the conservation law
for momenta which flow through the ladder (wavy lines in Fig. 2b);
the momenta are denoted as $q_1$ and $q_2$. Thus, the
phase space integration of the cut diagram is given by the
expression:
\be
d\Phi(P;k_1,k_2)d\Phi(P';k'_1,k'_2)(2\pi)^3
2k'_{10}\delta^{(3)}({\bf k}'_1-{\bf k}_1-{\bf
q}_1)(2\pi)^32k'_{20}\delta^{(3)}({\bf k}'_2 -{\bf k}_2+{\bf q}_2)\ .
\label{2b-3}
\ee
Here $k^2_i=k'^2_j=m^2$ and $P^2=s,P'^2=s'$.
We use light cone variables for the wave functions of the produced mesons;
the simplest way to introduce them is to use the infinite momentum frame.
Then the four-momenta $P\ =\left(P_0,{\bf P}_\perp,P_z\right)$
and  $k_i=\left(k_{i0},\  {\bf k}_{i\perp}\ ,k_{iz}\right)$
are written within the limits
$p\to \infty, k_{iz} \to \infty, k'_{iz} \to \infty$. After
introducing the light-cone variables ($\xi_i=k_{iz}/p$,
$\xi'_i=k'_{iz}/p$, $m^2_{i\perp}=m^2+k^2_{i\perp}$,
$m'^2_{i\perp}=m^2+k'^2_{i\perp}$) and putting
$q_{iz}$ small ($q'_{iz}/p\to 0$, the constraint of the multiperipheral
kinematics)
we re-write (\ref{2b-3}) as follows:
\begin{eqnarray}
&& \frac1{16\pi}\cdot\frac{d\xi_1d\xi_2}{\xi_1\xi_2}\delta(1-\xi_1-
\xi_2)d^2k_{1\perp}d^2k_{2\perp}\delta^{(2)}({\bf k}_{1\perp}+{\bf
k}_{2\perp})\delta\left(s-\frac{m^2_{1\perp}}{\xi_1}-
\frac{m^2_{2\perp}}{\xi_2}\right)\  \nonumber\\
&\times& \delta\left(s'+({\bf q}_{1\perp}-{\bf q}_{2\perp})^2-
\frac{m'^2_{1\perp}}{\xi_1}-\frac{m'^2_{2\perp}}{\xi_2}\right) \cdot
(2\pi)^32P_0\delta^{(2)}({\bf P}+{\bf q}_1-{\bf q}_2-{\bf P}')\
\nonumber\\
&\equiv & d\phi(\xi_1,\xi_2;{\bf k}_{1\perp},{\bf k}_{2\perp};{\bf
q}_{1\perp},{\bf q}_{2\perp})\cdot(2\pi)^32P_0\delta^{(3)}({\bf P} +
{\bf q}_1-{\bf q}_2-{\bf P}')\ .
\label{2b-4}
\end{eqnarray}
The factor $d\phi$ stands for the phase space integration in the
spectral integral (\ref{2b-1}): $d\phi \equiv
d\phi(\xi_1,\xi_2;{\bf k}_{1\perp},{\bf k}_{2\perp};{\bf
q}_{1\perp},{\bf q}_{2\perp})$.

In the approximation given by (\ref{2a-2}), when the jet block inside
of the big quark loop is considered as a real process, one may fix $q_1
=q_2 =0$; then the inclusive cross section is proportional  to
$T(p;0,0)$.
Taking $\delta$-functions into account in (\ref{2b-4}), the formula for
$T(p;0,0)$ acquires a rather simple form:
\be
T(p;0,0)\ =\ \frac1{16\pi^3}\int\limits^1_0
\frac{d\xi}{\xi(1-\xi)} \int d^2k_\perp\ \left(\frac{G_{\rm
meson}(s)}{s-\mu^2} \right)^2 S_{\rm meson}\ ,
\label{2b-5}
\ee
where $s=m^2_\perp/\left (\xi(1-\xi)\right )$.

The amplitude $T(p;0,0)$ alone does not determine the inclusive cross
section
$d\sigma /dx (Z^0\to {\rm meson} +X)$ because the
amplitudes $A(W^2_1,0)$ and $A(W^2_2,0)$ depend on $\xi$ and $k^2_\perp
$, see (\ref{2-2}). Taking into account this dependence, one has
at $x\sim 0$:
\be
\frac{d\sigma}{dx}(Z^0\to {\rm meson} +X)\ \sim \
 \frac1{16\pi^3}\int\limits^1_0
\frac{d\xi}{\xi(1-\xi)} \int d^2k_\perp\ \left(\frac{G_{\rm
meson}(s)}{s-\mu^2} \right)^2 S_{\rm meson}\ \Pi (W^2_1, W^2_2) \; .
\label{2b-6}
\ee
The spin factor
$S_{\rm meson}$ is closely related to the normalization of the wave
function of produced meson.

\subsection{Spin factors $S_\rho$ and $S_\pi$ }

Below, we calculate
the spin factors $S_\rho$ and $S_\pi$
at $q_1=q_2=0$. They are:
\begin{eqnarray}
&& S_\pi =\ -{\rm Sp}\left(i\gamma_5(\hat k_1+m)(\hat
k'_1+m)i\gamma_5(-\hat k'_2+m)(-\hat k_2+m)\right) \nonumber\\
&& S_\rho=\ -{\rm Sp}\left(\gamma^\perp_\alpha(\hat k_1+m)(\hat k'_1+m)
\gamma^\perp_\alpha(-\hat k'_2+m)(-\hat k_2+m) \right)\ .
\label{2s-1}
\end{eqnarray}
We have taken into account here that the quark--gluon ladder
carries quantum numbers of the scalar state, $J^P=0^+$; hence the
quark-ladder vertex is unity.

The $\rho$ meson vertex reads:
\be
\gamma^\perp_\alpha=g^\perp_{\alpha\alpha'}\gamma_{\alpha'}\ , \quad
g^\perp_{\alpha\alpha'}=g_{\alpha\alpha'}
-\frac{P_\alpha P_{\alpha'}}{P^2}\ .
\label{2s-2}
\ee
In the spin factor $S_\rho$ the summation is performed over
polarizations of the meson.
For the spin factors we have:
\be
S_\pi=8m^2s\ , \quad S_\rho=16\,m^2(s+2m^2)\ .
\label{2s-3}
\ee
Let us now demonstrate that similar spin factors determine the
normalization of wave functions of the $\rho$-meson and the pion.

\subsection{Wave functions of the pion and the $\rho$-meson}

In the framework of the light-cone technique
it is reasonable to introduce the wave
function of a particle and its normalization using the form factor of
the particle. The procedure of definition of the wave function
is discussed in detail in \cite{AMN,AMMP}. Schematically,
for the $q\bar q$ state this procedure looks as follows.

The form factor of a composite system (for definiteness, we consider the
pion form factor) is determined by the triangle diagram  Fig. 2c, where
the photon interacts with the composite system. The form factor is
represented as a double spectral integral over masses of the incoming
and outgoing pion; corresponding cuttings are
shown by the dashed lines I and II in Fig. 2c.

The structure of the amplitude of the triangle diagram for the pion has
the following form:
\be
A^{(tr)}_\nu\ =\ (p_\nu+p'_\nu)\, F_\pi(q^2)\ ,
\label{2wf-1}
\ee
where $p$ and $p'$ are momenta of the incoming and outgoing pions, the
index $\nu$ refers to the photon polarization and $F_\pi(q^2)$ is the pion
form factor which, in terms of the double spectral representation,
can be written as
\be
F_\pi(q^2)=\int\limits^\infty_{4m^2}\frac{dsds'}{\pi ^2}
\int d\phi^{(\rm tr)}(k_1,
k'_1,k_2)\frac{G_\pi(s)}{s-\mu^2}\,T_\pi(s,s',q^2)\ .
\label{2wf-2}
\ee
Here $d\phi^{(tr)}(k_1,k'_1.k_2)$ is the phase space of the triangle
diagram:
\be
d\Phi^{(\rm tr)}(k_1,k'_1,k_2)\ =\ d\Phi(P;k_1,k_2)d\Phi(P';k'_1,k'_2)
\cdot(2\pi)^3 2k'_{20}\delta^{(3)}({\bf k}'_2-{\bf k}_2)\ ;
\label{2wf-3}
\ee
$P$ and $P'$ are total momenta of the $q\bar q$ system before and
after the interaction with the photon: $P=k_1+k_2$ and $P'=k'_1+k'_2$, and
$P^2=s$ and $P'^2=s'$. The spin factor of the triangle diagram, $T_\pi$,
is determined by the following trace:
\be
(-){\rm Sp}\left[i\gamma_5(\hat k'_1+m)\gamma^\perp_\nu(\hat k_1+m)
i\gamma_5 (-\hat k_2+m)\right]=
\left(P^\perp_\nu+P'^\perp_\nu\right)T_\pi(s,s',q^2)\ .
\label{2wf-4}
\ee
The index $\perp$ stands for vectors orthogonal to the photon momentum:
\be
\gamma^\perp_\nu=\ g^\perp_{\nu\nu'}\gamma_{\nu'}, \quad
P^\perp_\nu=g^\perp_{\nu\nu'}P_{\nu'}, \quad g^\perp_{\nu\nu'}=
g_{\nu\nu'}-\frac{q_\nu q_{\nu'}}{q^2}\ .
\label{2wf-5}
\ee
At $q^2=0$ one has $F_\pi(0)=1$.
Direct calculations of equation (\ref{2wf-2}) in the limit $q^2\to 0$
give:
\be
1\ =\ \int\limits^\infty_{4m^2}
\frac{ds}\pi\left(\frac{G_\pi(s)}{ s-\mu^2}\right)^2 \rho(s)\,
S^{(wf)}_\pi(s)\ .
\label{2wf-6}
\ee
Here $\rho(s)$ is the phase volume
of the $q\bar q$ system:
\be
\rho(s)\ =\ \frac12\int d\Phi(P;k_1,k_2)\
=\ \frac1{16\pi} \sqrt{\frac{s-4m^2}s}\ ,
\label{2wf-7}
\ee
and $S_\pi^{(wf)}(s)$ is the trace of the quark loop diagram for the pion:
\be
S^{(wf)}_\pi (s)\ =\ (-){\rm Sp}\left[i\gamma_5(\hat k_1+m)i\gamma_5
(-\hat k_2+m)\right]\ =\ 2s\ .
\label{2wf-8}
\ee
Using light-cone variables we come to the following form of
(\ref{2wf-6}):
\be
1\ =\ \frac1{16\pi^3}\int\limits^1_0\frac{d\xi}{\xi(1-\xi)}\int
d^2{\bf k}_\perp\left(\frac{G_\pi(s)}{s-\mu^2}\right)^2 2s\ ,
\label{2wf-9}
\ee
where $s=(m^2+k^2_\perp)/\left (\xi(1-\xi)\right )$. This equation
enables us to introduce the pion wave function in the form
\be
\psi_\pi(\xi,{\bf k}_\perp)\ =\ \frac{G_\pi(s)}{s-\mu^2}\sqrt{2s}\ ,
\label{2wf-10}
\ee
which is normalized by the standard requirement.

Likewise,
we introduce the $\rho$-meson wave function:
it is defined by the
form factor which is the spin matrix $F_{\alpha \alpha'}(q^2)$.
In problems that do not
deal with polarization properties of the vector particle, it is convenient
to work with the trace of the form factor matrix,
$\sum_\alpha F_{\alpha \alpha}(q^2)$, which is normalized by
\be
\sum_{\alpha=1,2,3} F_{\alpha\alpha}(0)\ =\ 3\ .
\label{2wf-11}
\ee
The trace $\sum_\alpha F_{\alpha \alpha}(q^2)$
is determined by the expression analogous to (\ref{2wf-2}),
with evident substitutions $G_\pi \to G_\rho$ and
$T_\pi \to T_\rho$. As a
result, we obtain the normalization for averaged form factor:
\be
1\ =\ \frac13\sum_{\alpha=1,2,3} F_{\alpha\alpha}(0)\ =\
\int\limits^\infty_{4m^2}\frac{ds}\pi
\left(\frac{g_\rho(s)}{s-\mu^2}\right)^2 \rho(s) S^{(wf)}_\rho(s)\ ,
\label{2wf-12}
\ee
where
\be
S^{(wf)}_\rho(s)\ =\ -\frac13\ {\rm Sp}\left[\left(\gamma_\alpha
-P_\alpha\frac{\hat P}{P^2}\right)(\hat k_1+m)\left(\gamma_\alpha
-P_\alpha\frac{\hat P}{P^2}\right)(-\hat k_2+m)\right] .
\label{2wf-13}
\ee

The $\rho \to q\bar q$ vertex, $ \gamma_\alpha -P_\alpha
\hat P/P^2$, selects three degrees of freedom of the
$\rho$-meson. We have :
\be
S^{(wf)}_\rho(s)\ =\ \frac43 (s+2m^2)\ .
\label{2wf-14}
\ee
In the infinite momentum frame, (\ref{2wf-12}) is re-written as:
\be
1\ =\ \frac1{16\pi^2}\int\limits^1_0 \frac{d\xi}{\xi(1-\xi)}\int
d^2{\bf k}_\perp\psi^2_\rho(\xi,{\bf k}_\perp)\ ,
\label{2wf-15}
\ee
where
\be
\psi_\rho(\xi,{\bf k}_\perp)\ =\ \frac{G_\rho(s)}{s-\mu^2}
\sqrt{\frac43(s+2m^2)}\ .
\label{2wf-16}
\ee

\subsection{$V/P$ ratio for the central region and colour degrees of
freedom}

The normalization conditions for pion and $\rho$-meson wave
functions define unambigously the ratio of yields
for prompt production:
$\rho/\pi=3$, if the wave functions of these mesons are similar. Indeed,
the expression (\ref{2b-6}) represented in
terms of wave functions $\psi_{\rho}$ and $\psi_{\pi}$ give us
 (\ref{2-1}) immediately.

We have not taken into account explicitly the colour
degrees of freedom in the derivation.
This, however, can be easily done.
For the $meson \to q\bar q$ vertex the colour operator is equal to
$I/\sqrt{N_c}$,
where $I$ is a unity matrix in colour space. We have two colour
amplitudes, singlet and octet, for the
chain of the quark loop diagrams, $A_1$ and $A_8$.
The couplings of the amplitudes $A_1$ and $A_8$
 to quarks ($g(A_1)$ and $g(A_8)$) are proportional to
$I$ and $\blam$ (Gell--Mann matrices). All colour operators are the
same for both pion and $\rho$-meson production. Because of that, the
colour factors are not important for the ratio $\rho/\pi$ --- they are
identical and cancel in the production ratio.

As was stated above, the main contribution into inclusive meson
production comes from the ladder diagram $A_8$. This is due to the fact
that the coupling constant for the amplitude with $c=8$ is larger than for
$c=1$. In terms of $1/N_c$ expansion $g(A_1)/g(A_8) \sim 1/\sqrt{N_c}$.

\section{Inclusive production of mesons in the fragmentation region}

The inclusive production
cross section of mesons in the fragmentation region is
determined by the discontinuity of the diagram 2d. The spectral
representation for this diagram is written as an integral over
the masses of initial and final $q\bar q$ states
in the transitions $Z^0\to q\bar q$ and $q\bar q\to Z^0$ and
over $q\bar q$ masses in the transitions
$q\bar q\to meson$ and $meson\to q\bar q$.
The amplitude of the
diagram of Fig. 2d reads:
\begin{eqnarray}
&& \int \limits_{4m^2}^{\infty}
\frac{dM^2dM'^2}{\pi^2}\ \frac{g_Z(M^2)g_Z(M'^2)}{(M^2-M^2_Z)
(M'^2-M^2_Z)}\int \limits_{4m^2}^{\infty}
\frac{dsds'}{\pi^2}\psi_{\rm meson}(s)\psi_{\rm
meson}(s')  \nonumber\\
&\times&\;  d\phi_3(k_1,k_2,k_3)d\phi_3(k'_1,k'_2,k'_3)
A(W^2,(k_2-k'_2)^2)
\frac{S^{(fr)}_{\rm meson} }{\sqrt{S^{(wf)}_{\rm meson}(s)S^{(wf)}_{\rm
meson}(s')}}\ .
\label{3-1}
\end{eqnarray}
The vertices $g_Z(M^2)$ and $g_Z(M'^2)$ are written
for the $Z^0\to q\bar q$ and $q\bar q\to Z^0$ transitions.
Spectral integrals over $s$ and $s'$
stand for $q\bar q\to meson$ and $meson\to q\bar q$
(where $meson$ means $\pi,\rho$).
The wave function
$\psi_{\rm meson}$ of the produced meson was introduced in an
explicit form in Section 2 for the pion and the $\rho$-meson, and the
factors $d\phi_3(k_1,k_2,k_3)$ and
$d\phi_3(k'_1,k'_2,k'_3)$ define the integration over phase spaces
in the left- and right-hand parts of the diagram of Fig. 2d:
\begin{eqnarray}
&& d\phi_3(k_1,k_2,k_3)= \frac12\, \frac{d^3k_1}{(2\pi)^3 2k_{10}}
\frac{d^3k_2}{(2\pi)^32k_{20}}\ ,  \nonumber\\
&&\times (2\pi)^4\delta^{(4)}\left(\tilde
P-k_1-k_2\right)\  \frac12\, \frac{d^3k_3}{(2\pi)^32k_{30}}\,
(2\pi)^4\delta^{(4)}(P-k_1-k_3) \ .
\label{3-2}
\end{eqnarray}
Here $\tilde P^2=M^2$ and $ P^2=s$.

The block $A\left (W^2,
(k_2-k_2')\right ) $ defines the multiperipheral ladder
(wavy line in Fig.
2d). This block depends on the
momentum transfer squared $(k_2-k'_2)^2$ and the total energy squared
$W^2$:
\be
W^2\ \simeq\ M^2_Z(1-x)\ ;
\label{3-3}
\ee
$x$ is the momentum fraction carried by the
produced meson: $x=2p/M_z$,
where $p$ is the longitudinal component of meson momentum,
$p_{\rm meson}=(p+\mu^2_\perp/2p, 0,p)$.

The spectra $d\sigma(Z^0\to meson+X)/dx$ fall rapidly
with increasing $x$: this decrease is governed by
$A\left (W^2,(k_2-k'_2)\right )$.

All the characteristics of (\ref{3-1}) listed above are the same for
both pion and $\rho$-meson production, the wave functions $\psi_{\pi}$
and $\psi_{\rho}$ are also supposed to be the same.  The difference may
be contained in the spin factors $S^{(fr)}_\pi$ and $S^{(fr)}_\rho$
which are
\begin{eqnarray} S^{(fr)}_\pi &=&(-){\rm
 Sp}\bigg[\gamma'^\perp_\nu(1+R\gamma_5)(\hat k'_1+m)i\gamma_5
(\hat k'_3+m) (\hat k_3+m) \nonumber\\
&\times &\;  i\gamma_5(\hat
k_1+m)\gamma^\perp_\nu (1+R\gamma_5)(-\hat k_2+m)(-\hat k'_2+m)\bigg];
\nonumber\\
S^{(fr)}_\rho&=&(-){\rm
 Sp}\bigg[\gamma'^\perp_\nu(1+R\gamma_5)( \hat
k'_1+m)\gamma'^\perp_\alpha(\hat k'_3+m)(\hat k_3+m)
\nonumber\\
&\times &\; \gamma^\perp_\alpha
(\hat k_1+m)\gamma^\perp_\nu(1+R\gamma_5)(-\hat k_2+m)(-\hat
k'_2+m)\bigg].
\label{3-4}
\end{eqnarray}
The factor $\gamma_\nu^\perp (1+R\gamma_5)$ is related to the vertex
$Z^0 \to q\bar q$ which is determined by the vector and axial--vector
interactions (the ratio of coupling constants is
 $\sim 2.63$ for the $u$ quark
and $\sim 1.43$ for the $d$ quark).
For $S^{(fr)}_\pi$ the summation is performed over
polarizations of the $Z^0$ boson (index $\nu$); $\gamma_\nu^\perp $
is orthogonal to $k_1+k_2$ and $\gamma'^\perp_\nu $ is orthogonal to
$k'_1+k'_2$.  For the $\rho$-meson
spin factor $S^{(fr)}_\rho$, the
summation is performed over polarizations of the $Z^0$ boson (index
$\nu$) and $\rho$-meson (index $\alpha$) with
 $\gamma_\alpha^\perp \; \perp \; (k_1+k_3)$ and $\gamma'^\perp_\alpha
\; \perp \; (k'_1+k'_3)$ (see also (\ref{2s-2})).

Normalization factors $S^{(wf)}_{\rm meson}(s)$ and $S^{(wf)}_{\rm
meson}(s')$ in (\ref{3-1}) are related to the definition of meson
wave functions, see (\ref{2wf-10}) and (\ref{2wf-16}).
The explicit expressions for spin factors (\ref{3-4}) are presented
in Appendix B.
Using these expressions, we have an opportunity to perform
numerical evaluation of ratios of prompt meson yields in the decay
reactions with smaller energy release. In the case of the $Z^0$
decay, when $M_Z >> m$ and $M_Z >> \mu$, a reasonably good
approximation  for (\ref{3-1}) is
to substitute the integrals over $M^2$ and
$M'^2$ by semi-residues in the poles $M^2=M^2_z$ and $M'^2=M^2_z$.
Then we have the following kinematics for real jets:
\be
k_1=k'_1\ , \quad k_2=k'_2\  .
\label{3-6}
\ee
In this approximation we can put $ k_3=k'_3$ and $s=s'$.
Then we have for light mesons:
\be
\frac{S^{(fr)}_\pi}{S^{(wf)}_\pi}\simeq2M^2_Z(1+R^2)\ , \quad
\frac{S^{(fr)}_\rho}{S^{(wf)}_\rho}\simeq6M^2_Z(1+R^2)\ ,
\label{3-7}
\ee
which gives
the ratio of the prompt yields $\rho:\pi=3:1$ in the
fragmentation region $x\sim 0.5-1$ (more generally, $V:P=3:1$ for
hadronic decays $Z^0\to q\bar q$ with $q=u,d,s$).

The same ratio appears
for the production of heavy quarks $Z^0=Q\bar Q$, where $Q=c,b$.
For example, in the case of $b$ quark the spin factors are:
\begin{eqnarray}
&& \frac{S^{(fr)}_B}{S^{(wf)}_B}\ \simeq\
2\left[M^2_Z+2m^2_b+R^2(M^2_Z-4m^2_b)\right]\ , \nonumber\\
&& \frac{S^{(fr)}_{B^*}}{S^{(wf)}_{B^*}}\ =\ 6\left[M_Z^2+2m^2_b+R^2(
M^2_Z-4m^2_b)\right]\ ,
\label{3-8}
\end{eqnarray}
and thus the ratio $B^*_{prompt}:B_{prompt}$ also equals 3.

\subsection{$V/P$ ratio in the fragmentation region: comparison to data}

We have seen that $V/P=3$ in the fragmentation region as well as in
the central
region. However, in the central region the comparison of quark
combinatorics with experiment is hampered by the presence of a number
of the decay products of highly excited resonances, while in the
fragmentation region this contribution is suppressed by rapidly
decreasing spectra. This means that the fragmentation region allows us
to perform a model--independent verification of quark combinatorics.
Here we carry out such a comparison based on the ALEPH data \cite{ALEPH}.

To operate with meson spectra at $x\sim 0.2-0.8 $ we have fitted the
spectra $(1/\sigma_{tot})d\sigma/dx$ to the sum of exponents $\sum
C_ie^{-b_ix}$; the calculation results are presented in Fig. 3 for
$\pi^\pm, \pi^0, \rho^0$ and $(p,\bar p)$. The ratio of fitting curves
drawn with calculation errors (shaded area) is shown in Fig. 4a for
$\rho/\pi$. We see that for $0.6 < x <0.8$ the data are in  reasonable
agreement with the prediction $\rho/\pi=3$.

The figure 4b,c demonstrates the ratios $K^{*0}/K^0$ and
 $K^{*\pm}/K^\pm$:  the data do not contradict the prediction, though
the errors are too large to conclude anything more definite.

\section{Suppression of the strange and heavy quark productions}

In hadronic multiparticle production processes (in jet processes of
the type of $Z^0\to hadrons$ or in hadron--hadron
collisions at high energies) the production of strange quarks is
suppressed. Strong suppression is observed for the
production of heavy quarks $Q=c,b$. One can guess that this
suppression, being of the same nature for different reactions, is
related to the mechanism of the production of new quarks at large
separations of colour objects. This mechanism is seen in its pure
form in the two--particle decays (the corresponding diagram is shown in
Fig.  5a).  The block of the production of a new $q\bar q$ pair in the
two--particle  decay is the same as that of meson production
in jet processes (Fig. 5b), so it is reasonable to suppose that the
suppression mechanism of the production of new quarks is similar
for these processes.

\subsection{Two--meson decay of the $q\bar q$ state
and soft hadronization}

The decay of the $q\bar q$ state takes place
as follows: the quarks of the excited state leave
the region where they were kept by
the confinement barrier, and at a sufficiently large separation
a new quark--antiquark pair will be produced inevitably: together with
the incident quarks, these new quarks then form mesons (i.e. free
particles).
Schematically, this process (which is the leading one in terms
of the $1/N$ expansion)
is shown on the diagram of Fig. 5a:
two quarks fly away (with the momenta $p_1$ and $p_2$), and at large
quark separations the gluonic field produces a new $q\bar q$ pair (the
quarks with momenta $k_2$, $k_3$); then the primary quark (now with
momentum $k_1$) joins the newly-born one ($k_2$) creating a meson.
Likewise,  another newly-born quark ($k_3$) joins the other primary
quark (now with momentum $k_4$) producing the second meson.

The block with the quark-antiquark production, that is the transition
\be
q(p_1)+\bar q(p_2) \to q(k_1)+\bar q(k_2)+q(k_3)+\bar q(k_4)
\label{1rep}
\ee
is the key process that determines the decay physics; it is
shown separately
in Fig. 5c.
The process (\ref{1rep}) is responsible for the leaving the
confinement trap by quarks.
In modelling the hadronization transition amplitude
(\ref{1rep}),
quark combinatorics uses the hypothesis of soft
hadronization.  The idea was formulated at the 70's, and even now
the soft hadronization hypothesis looks rather reliable and productive.
It suggests that in the ladder
of produced quarks (Figs. 1a, 5c) the
contribution comes from
small momentum transfers
(of the order of $R^{-2}_{confinement}$).
In the framework of the space--time picture this means that
new $q\bar q$ pairs are produced at large separations,
when colour objects leave the confinement well.

The soft hadronization hypothesis applied to the decay processes treats
the ladder diagram of Fig. 5c for
the decay amplitude of Fig. 5a in the same way as for
jet production, Fig. 5b:
process (\ref{1rep}) is an elementary subprocess both for the
high--energy ladder and the two--particle decay amplitude, and
the momentum transfers which enter the
amplitude (\ref{1rep}) appear to be small in the hadronic scale,
$\sim R^{-2}_{confinement}$.

\subsection{Spectral representation of the decay diagram}

Below we consider in detail the decay amplitude of Fig. 5a,
performing calculations, as is done before, in the framework of the
spectral representation with the light-cone wave functions for
$q\bar q$ states.

By using the notations
\begin{eqnarray}
&& P\ =p_1+p_2, \quad k_{12}\ =k_1+k_2, \quad k_{34}\ =k_3+k_4,
 \nonumber\\
&& M^2\ =(k_1+k_2)^2, \quad s_{12}\ =(k_1+k_2)^2, \quad
s_{34}\ =(k_3+k_4)^2
\label{3rep}
\end{eqnarray}
we have the following spectral representation for the
amplitude of the diagram shown in Fig. 5a:
\begin{eqnarray}
&&A(q\bar q \,\; {\rm state} \to {\rm two} \, {\rm mesons})=
\int \limits_{4m^2}^{\infty} \frac {dM^2}{\pi}
\Psi_{in}(M^2) d\Phi(P;p_1,p_2) \int \limits_{(m+m_s)}^\infty
\frac {ds_{12}ds_{34}}{\pi^2}     \nonumber \\
&&\times \; t(p_1,p_2;k_1,k_2,k_3,k_4)d\Phi(k_{12};k_1,k_2)\;
d\Phi(k_{34};k_3,k_4)
\Psi_1(s_{12}) \Psi_2(s_{34})\ .
\label{4rep}
\end{eqnarray}
Here the masses of newly--born quarks $i=2,3$ are denoted as $m_s$,
thus opening the way to consider the decay with strange quark production.
The transition  amplitude (\ref {1rep}) of Fig. 5c
is denoted as  $t(p_1,p_2;k_1,k_2,k_3,k_4)$.
The decay amplitude (\ref {4rep}) is written in
terms of meson wave functions: for the initial state it is
$\Psi_{in}(M^2)$, and for outgoing mesons they are
$\Psi_1(s_{12})$ and $\Psi_2(s_{12})$.  We do not specify here the
meson spin structure of the quark propagator:  we consider it below in
a more detailed consideration of the transition amplitude
(\ref{1rep}).

Thus, the decay amplitude $A$ is a convolution of the transition amplitude
(\ref{1rep}) with wave functions of initial and outgoing mesons:
\be
A(q\bar q \,\; {\rm state} \to {\rm two} \, {\rm mesons})=
\Psi_{in}\otimes t \otimes \Psi_1 \Psi_2\ .
\label{6rep}
\ee
Further, we use
light-cone variables;
with $p_{iz}\to \infty$ and $k_{iz}\to \infty$
we have
\bea
&&k_i=(k_{iz}+\frac{m^2+k_{i\perp}^2}{2k_{iz}},
\bk_{i\perp},k_{iz}) \qquad i=1,4 \nonumber \\
&&k_i=(k_{iz}+\frac{m^2_s+k_{i\perp}^2}{2k_{iz}},
\bk_{i\perp},k_{iz}) \qquad i=2,3 \nonumber \\
&& p_1=(p_{iz}+\frac{m^2+p_{i\perp}^2}{2p_{iz}},
\bp_{i\perp},p_{iz}), \qquad i=1,2.
\label{7rep}
\eea
Let us use the frame where the outgoing particles
are moving along the $z$-axis.
We put
\bea
&& P=(P_z+\frac{M^2}{2P_z}, 0,P_z),   \nonumber \\
&& k_{12}=(k_{12z}+\frac{s_{12}}{2k_{12z}}, 0,k_{12z}), \nonumber \\
&& k_{34}=(k_{34z}+\frac{s_{34}}{2k_{34z}}, 0,k_{34z}).
\label{9rep}
\eea
The phase spaces $d\Phi(k_{12};k_1,k_2)$,
$d\Phi(k_{34};k_3,k_4)$ contain the
energy-momentum conservation $\delta$-functions which give
\bea
&& M^2=\frac{m^2+p^2_{1\perp}}{x_1} +\frac{m^2+p^2_{2\perp}}{x_2},
\nonumber \\
&& s_{12}=\frac{m^2+k^2_{1\perp}}{y_1} +\frac{m^2_s+k^2_{2\perp}}{y_2} ,
\nonumber \\
&& s_{34}=\frac{m^2_s+k^2_{3\perp}}{y_3} +\frac{m^2+k^2_{4\perp}}{y_4}
.
\eea
Here
$x_i=p_{iz}/(p_{1z}+p_{2z})$ , $y_i=k_{iz}/(k_{1z}+k_{2z})$ for
$i=1,2$ and $y_i=k_{iz}/(k_{3z}+k_{4z})$ for $i=3,4$.

\subsection{The region of dominance of the transition amplitude
$t_{q\bar q \to q\bar q q\bar q}$}

Let us turn to the
principal point, namely: the
evaluation of the region in momentum space
selected by the transition amplitude of Fig. 5c within the soft
hadronization assumption.

The hypothesis of soft hadronization means that the ladder
diagram of Fig. 5c
has a peripheral structure: it requires
the momentum transfers squared to the $q\bar q$ block to be
small, of the order of $1/R^2_{confinement}\;$. So
\bea
&&-(p_1-k_1)^2 \simeq (\bp_{1\perp}-\bk_{1\perp})^2 \sim
R^{-2}_{confinement}\; , \nonumber\\
&&-(p_2-k_4)^2 \simeq (\bp_{2\perp}-\bk_{4\perp})^2
\sim R^{-2}_{confinement}\; .
\label{11rep}
\eea
Likewise, the momentum transfer squared in the quark propagator (see
Fig. 5c) is of the order:
\be
(p_1-k_1-k_2)^2 \simeq -(\bp_{1\perp}-\bk_{1\perp}-
\bk_{2\perp})^2.
\ee

Let us now discuss the peripheral constraint (\ref{11rep}) in  more
detail.  As an example, let us consider $(p_1-k_1)^2$. We have
\bea
&&(p_1-k_1)^2=(p_{1z}-k_{1z}+\frac{m^2+p_{1\perp}^2}{2p_{1z}}-
\frac{m^2+k_{1\perp}^2}{2k_{1z}})^2\ ,
\nonumber \\
&&-(\bp_{1\perp}-\bk_{1\perp})^2-(p_{1z}-k_{1z})^2\ .
\eea
The gluon carries a small fraction of the longitudinal momentum of
the primary quark, therefore
\be
p_{1z} >> p_{1z} -k_{1z} \equiv \Delta.
\ee
Then
\be
(k_1-p_1)^2 \simeq
\frac{\Delta}{p_{1z}}(k_{1\perp}^2-p_{1\perp}^2)-
(\bk_{1\perp}-\bp_{1\perp})^2
\simeq -(\bk_\perp-\bp_{1\perp})^2  \ .
\ee
The peripheral constraint means that $y_1>>y_2$.
But in numerical calculations this does not require
very large invariant energies squared, $s_{12}$  and $s_{34}$, which
determine the blocks of quark fusion. For example,
at $y_2/y_1 \sim 1/10$ one has $s_{12} \sim 10 m^2 \sim 1$ GeV$^2$
for $m\simeq 350 $ MeV, which gives for the quark relative momentum
the value of the order of $300 -400$ MeV.

\subsection{The decay amplitude
$A(q\bar q \, {\rm state} \to {\rm two} \, {\rm mesons})$}

Let us now write the formula for $t(p_1,p_2;k_1,k_2,k_3,k_4) $
in the simplest approximation, taking into account the
$t$-channel propagators only:
\bea
&&t(p_1,p_2;k_1,k_2,k_3,k_4)=\frac{g}{m^2_g+(\bp_{1\perp}-\bk_{1\perp})^2}
\nonumber \\
&&\times  \frac{g^2
\left (m_s-\bgam(\bp_{1\perp}-\bk_{1\perp}-\bk_{2\perp})\right  )}
{m^2_s+(\bp_{1\perp}-\bk_{1\perp}- \bk_{2\perp})^2} \cdot
\frac{g}{m^2_g+( \bp_{2\perp}-\bk_{4\perp})^2}
\label{16rep}
\eea
To avoid ultrared divergence, the effective mass of the soft gluon is
introduced into the gluon propagator  (see, for example,  \cite{L}).

The equation (\ref{16rep}) does not state that the transition
amplitude $t(p_1,p_2;k_1,k_2,k_3,k_4)$
selects large distances. At this point the amplitude (\ref{16rep}) may
be improved by incorporating form factors into the gluon emission
vertex:
\be
g\to g((\bp_{1\perp}-\bk_{1\perp})^2), \quad
g\to g((\bp_{2\perp}-\bk_{4\perp})^2)\ .
\label{17rep}
\ee
With equations (\ref{16rep}) and (\ref{17rep}) the amplitude $A$ reads:
\bea
&&A(q\bar q \,\; {\rm state} \to {\rm two} \, {\rm mesons})
=\int \limits_0^1
\frac{dx_1dx_2\delta(1-x_1-x_2)}{16\pi^2 x_1x_2}
d\bp_{1\perp}d\bp_{2\perp}
\delta(\bp_{1\perp}+\bp_{1\perp})
\nonumber \\
&&\times \int \limits_{y_1\gg y_2}
\frac{dy_1dy_2\delta(1-y_1-y_2)}{16\pi^2 y_1y_2}\cdot
d\bk_{1\perp}d\bk_{2\perp}
\delta(\bk_{1\perp}+\bk_{2\perp})
\nonumber \\
&&\times\int \limits_{y_4\gg y_3}
\frac{dy_3dy_4\delta(1-y_3-y_4)}{16\pi^2 y_3y_4}
d\bk_{3\perp}d\bk_{4\perp}
\delta(\bk_{3\perp}+\bk_{4\perp})
\Psi_{in}(x_1,x_2;\bp_{1\perp},\bp_{2\perp})
\nonumber \\
&&\times \frac{g ((\bp_{1\perp}-\bk_{1\perp})^2 )}
{m^2_g+(\bp_{1\perp}-\bk_{1\perp})^2}
\frac{g^2 (m_s- \bgam (\bp_{1\perp}-\bk_{1\perp}-\bk_{2\perp})) }
{m^2_s +(\bp_{1\perp}-\bk_{1\perp}-\bk_{2\perp})^2 }
\frac{g ((\bp_{2\perp }-\bk_{4\perp })^2 ) } {m^2_g+(\bp_{2\perp}
-\bk_{4\perp })^2 }
\nonumber \\
&&\times \Psi_1(y_1,y_2;\bk_{1\perp},\bk_{2\perp})
\Psi_2(y_3,y_4;\bk_{3\perp},
\bk_{4\perp}) \ .
\label{18rep}
\eea
This is the expression for the decay amplitude which makes it
possible to discuss the rules of quark combinatorics.

In a rough approximation that still
gives a qualitatively correct answer, we  neglect the
momenta in the propagator of newly--born quarks:
\be
\frac{g^2\left (m_s-\bgam
(\bp_{1\perp}-\bk_{1\perp}-\bk_{2\perp})\right )}{m_s^2+
(\bp_{1\perp}-\bk_{1\perp}-\bk_{2\perp})^2} \to \frac{g^2}{m_s}\ .
\label{19rep}
\ee
We have
\be
A(q\bar q \,\; {\rm state} \to {\rm two} \, {\rm mesons})
=\frac{\alpha_s}{m_s}
\cdot \left (
\Psi_{in}\otimes t \otimes \Psi_1\Psi_2
\right )\ ,
\label{20rep}
\ee
This equation tells us
that the probability to produce non-strange and strange
quarks, $u\bar u:d\bar d:s\bar s=1:1:\lambda$,
is determined by the ratio of masses squared of
non-strange $(u,d)$ to strange $(s)$ quark. Introducing the
constituent quark masses in the soft region, $m_u\simeq
m_d\equiv m = 350$ MeV and $m_s \simeq 500$ MeV, we get:
\be
\lambda\simeq \frac{m^2}{m^2_s}\simeq 0.5\ .
\label{21rep}
\ee
The equations (\ref{19rep}) and (\ref{20rep})
justify the statements
of  quark combinatorics applied to the decay processes
\cite{4,5,6,Peters}. Of course,
we here suppose the identity of
meson wave functions belonging to the same multiplet.

The equation (\ref{21rep}) gives us a rough evaluation of
$\lambda$, for in (\ref{19rep}) we neglected
momentum transfers squared that are compatible with light quark
masses. In
more sophisticated evaluations of $\lambda$,
one may
take into account the momentum dependence of the quark propagator:
\be
\frac{1}{m_s^2+
(\bp_{1\perp}-\bk_{1\perp}-\bk_{2\perp})^2} \to
\frac1{m_s^2 +<k^2>}\ ,
\ee
where $<k^2>$ is a typical momentum squared
inherent to the considered decay process.
Therefore,
\be
\lambda =\frac{m^2+<k^2>}{m_s^2+<k^2>} \; .
\label{26rep}
\ee
For standard
decays of light resonances $<k^2>\sim 0.1-0.3$ (GeV/c)$^2$,
and this results in the increase of $\lambda$ compared to the
estimation (\ref{21rep}).  Indeed, in the analysis \cite{Peters}
$\lambda \sim 0.7$ was found.  Actually, equation (\ref{26rep})
demonstrates that $\lambda$ can vary depending on different types of
reactions.

\subsection{Suppression parameter $\lambda$ in the hadronic $Z^0$
decay }

Now let us turn to the processes of the $q\bar q$ pair production in
jet processes $Z^0 \to q\bar q \to hadrons$, which is shown in Fig. 5b.
All the above considerations, which have been used for the decay of a
resonance into two mesons, may be applied to this process. As a result,
we obtain the following formula which is
a counterpart of (\ref{20rep}):
\be
A(Z^0 \to {\rm two}\, {\rm mesons}+X)=\frac{\alpha_s}{m_s}
\cdot \left (q\bar q \,{\rm from}\,{\rm jet}\,{\rm ladder}
\otimes t \otimes \Psi_1\Psi_2\right )\ .
\label{20rep1}
\ee
This formula differs from (\ref{20rep}) by the initial state only,
which is the wave function of $q\bar q$-pair
for  jet ladder but not for the
state defined by the wave function $\Psi_{in}$. This means that
the ratio of
probabilities of producing a strange and a non-strange quark are given by
the factor $m^2/m^2_s$. So, like in the decay process, one has:
\be
\lambda\simeq\frac{m^2}{m^2_s}\simeq 0.5\ .
\ee
Experimental data on the ratio of yields $K^{\pm}/\pi^\pm$ do not
contradict this evaluation. The ratio  $K^{\pm}/\pi^\pm$ as a function
of $x$ is shown in Fig. 4e. It is seen that at $x=0.2$
$K^{\pm}/\pi^\pm\simeq 0.35$. With the increase of $x$ the ratio
$K^{\pm}/\pi^\pm$ grows and reaches the value $\sim 0.8$ at $x =0.7$.
Such an increase is rather legible: the matter is that $K$ mesons are
produced both due to the formation of a new $s\bar s$ pair in the ladder
(with the probability $\lambda$) and fragmentation production of $s\bar
s$ pair in the transition $Z^0 \to s\bar s$. Relative probabilities of
prompt production of $Z^0 \to
u\bar u,d\bar d,s\bar s$  obey the ratio
$u\bar u:d\bar d:s\bar s\simeq 0.26:0.37:0.37$. Because of that the
production of $K$ meson at large $x$ is proportional to
\be
K^+ \sim (0.37\cdot 1 +0.26\cdot \lambda)\ .
\label{28rep}
\ee
Let us comment this formula in a more
detail. The $K^+$ meson is produced in the two processes: (i)
fragmentation production of $\bar s$ quark (relative probability
0.37) and subsequent pick-up of the $u$ quark from $q\bar q$ sea in a
jet (relative probability 1), and (ii) fragmentation production of $u$
quark (relative probability 0.26) and pick-up of $\bar s$ quark from
the sea (relative probability $\lambda$).

The same quantity for pions is
\be
\pi^+ \sim (0.37\cdot 1 +0.26\cdot 1)\ .
\label{29rep}
\ee
So, the ratio $K^+/\pi^+$ at large $x$ is equal to
\be
\frac{K^+}{\pi^+}=\frac{0.37+0.26\lambda} {0.37+0.26}\simeq 0.8\
\label{30rep}
\ee
at $\lambda=0.5$.
This value agrees with the experimental data
\cite{20} as it is demonstrated by Fig. 4e,
where $\lambda(x)$ determined as the $ K^{\pm}/\pi^{\pm}(x)$-ratio is
shown.

The small value of $K^+/\pi^+$ at $x=0$
is a direct consequence of large probability to produce
highly excited resonances: in the resonance decay more pions than
kaons are produced. For the problem of
breeding of strange and non-strange states in the decay, it is
rather interesting to see the ratio $K^*/\rho$ --- experimental data
for  $K^{0*}/\rho^0$ are shown in Fig. 4e too (shaded area).
Remarkably, the ratio  $K^{0*}/\rho^0$  has no tendency to
decrease with decreasing $x$: this means that the rate of
breeding of $K^{0*}$ and $\rho^0$ in decays is approximately the same.
Unfortunately, experimental errors are too large to have  more
definite conclusions about the behaviour of $\lambda(x)$.

\subsection{Production of heavy quarks}

The supression parameter for the production of a strange quark cannot be
reliably determined. This is because the masses of strange
and non-strange quarks are small compared to the mean
transverse momenta of quarks in the production process, see
(\ref{26rep}).
We can draw a more
definite conclusion
about the
suppression parameter $\lambda_Q$ for the production of heavy quarks
$Q=c,b$. This parameter is defined by the same formula (\ref{18rep})
for multiperipheral production, so we have:
\be \lambda_Q \simeq
\frac{m^2}{m^2_Q\ln^2\frac{\Lambda^2}{m^2_Q}}.
\ee
Here we take into account that the gluon--quark
coupling constant decreases with the
growth of the quark mass, $\lambda_Q\sim \alpha^2(m^2_Q)$. The QCD scale
constant, $\Lambda$, is of the order of 200 MeV.

To estimate $\lambda_c$ and $\lambda_b$, let us use
$m=0.35$ GeV, $m_c=M_{J/\Psi}/2=1.55$ GeV, $m_b=M_{\Upsilon}/2=4.73$
GeV and $\Lambda=0.2$ GeV. Then
\be
\lambda_c \simeq 2.8\cdot 10^{-3}, \qquad
\lambda_b \simeq 1.1\cdot 10^{-4}.
\label{heavy}
\ee
The value of $\lambda_c $ should reveal itself in the
inclusive production of $J/\Psi$ and $\chi$ mesons, while $\lambda_b$
is to be seen
in reactions with $\Upsilon$'s: $Z^0 \to (\sum J/\Psi + \sum
\chi)+X$ and $Z^0 \to \sum  \Upsilon +X$.
These reactions are
determined by the processes
$Z^0 \to c\bar c \to c+(\bar c c +\bar q q{\rm -sea})+\bar c$  and
$Z^0 \to b\bar b \to b+(\bar b b +\bar q q{\rm -sea})+\bar b $: for the
production of a $c\bar c$ or $b\bar b$ meson a new pair of heavy quarks
should be produced, since the quarks formed at the first stage of the
decay, $Z^0 \to c\bar c $ or $Z^0 \to b\bar b $, have a rather big gap
in the rapidity scale. So, within the definition \bea &&\lambda_c
\simeq \Gamma
\bigg (c\bar c \to c+(\bar c c +\bar q q{\rm -sea} )+\bar c \bigg )
/\Gamma (c\bar c)
\nonumber \\
&&\lambda_b \simeq
\Gamma \bigg  (b\bar b \to b+(\bar b b +\bar q q{\rm -sea} )+
\bar b\bigg )
/\Gamma (b\bar b),
\eea
we estimate $ \Gamma (c\bar c \to c+(\bar c c +\bar q q{\rm -sea} )
+\bar c)$
and $\Gamma (b\bar b \to b+(\bar b b +\bar q
q{\rm -sea} )+\bar b)$
by
the available data from \cite{PDG}:
\bea
&&\Gamma
\bigg  (c\bar c \to c+(\bar c c +\bar q
q{\rm -sea} )+\bar c \bigg  ) \sim
\Gamma
\bigg  (J/\Psi(1S)X) +\Gamma (J/\Psi(2S)X) +\Gamma (\chi(1P)X \bigg  ),
\nonumber \\
&&\Gamma
\bigg (b\bar b \to b+(\bar b b +\bar q
q{\rm -sea} )+\bar b\bigg  ) \sim
\Gamma \bigg  (\Upsilon(1S)X+ \Upsilon(2S)X+\Upsilon(3S)X \bigg  ).
\eea
Experiment gives \cite{PDG}:
\bea
&&\Gamma (c\bar c)/\Gamma (hadrons)= 0.177\pm 0.008,
\nonumber \\
&&\Gamma (b\bar b)/\Gamma (hadrons)= 0.217\pm 0.001,
\nonumber \\
&&\Gamma\bigg  (J/\Psi(1S)X +J/\Psi(2S)X +\chi(1P)X\bigg )
/\Gamma (hadrons) =(1.17\pm 0.13)\cdot 10^{-2} ,
\nonumber \\
&&\Gamma \bigg (\Upsilon(1S)X+ \Upsilon(2S)X+\Upsilon(3S)X\bigg )
/\Gamma (hadrons)
=(1.4\pm 0.9)\cdot 10^{-4}\ .
\eea
Thus, we have
\bea
&&\Gamma \bigg (J/\Psi(1S)X +J/\Psi(2S)X +\chi(1P)X\bigg )
/\Gamma (c\bar c)
=(2.07\pm 0.23)\cdot 10^{-3},
\nonumber \\
&&\Gamma \bigg (\Upsilon(1S)X+ \Upsilon(2S)X+\Upsilon(3S)X\bigg )
/\Gamma (b\bar b)
=(0.31 \pm 0.19)\cdot 10^{-4}\ ,
\eea
in a reasonable agreement with (\ref{heavy}).

\section{The problem of saturation of the produced \\
$q\bar q$ and $qqq$ states by mesons and baryons; \\
baryon-meson ratio and the Watson-Migdal factor }

In quark combinatorics formulated in \cite{1,2} the baryon quark number
reveals itself as the probability to produce a baryon containing this
quark.
For the hadronization of the quark $q_i$ in $(q,\bar q)$-sea
the production rule reads:
\begin{equation}
q_i+(q,\bar q)_{\mbox {sea}} \to \frac1{3}B_i+\frac{2}{3}M_i+
\frac1{3}M+(M,B,\bar B)_{\mbox {sea}},
\label{2-hep}
\end{equation}
where $B_i$ and $M_i$ are baryons and mesons containing the quark $q_i$
and $M$, $B$ and $\bar B$ are mesons, baryons and antibaryons of the sea.

Generally one can write
\begin{equation}
M\ =\ \sum_L\mu_L M_L \; ,\qquad  B\ =\
\sum_L\beta_LM_L\ ,
\label{3-hep}
\end{equation}
$$
M_i\ =\ \sum_L\mu_L^{(i)}  M_L^{(i)} \; ,\qquad  B_i\ =\
\sum_L\beta_L^{(i)} M_L^{(i)} \ .
$$
where indices
$L=0,1,2,...$ define the multiplet, while $\mu_L ,\; \mu_L^{(i)}$ and
$\beta_L ,\; \beta_L^{(i)} $
are production probabilities of mesons and baryons of the given
multiplet in the process of quark hadronization. These
probabilities are determined by characteristic relative momenta
of the fused quarks.

\subsection{The baryon-meson ratio}

Our present understanding of the multiperipheral ladder is not sufficient
to re-analyse
(\ref{2-hep}) on the level carried out
in Sections 2 and 3 for $V/P$.
Nevertheless, the data for decays $Z^0\to p+X$ and $Z^0\to \pi^++X$
definitely confirm the equation (\ref{2-hep}). Quark combinatorics \cite{1*}
predict for $p/\pi^+$  at large $x$:
\begin{equation}
p/\pi^+\simeq 0.20
\label{5-hep}
\end{equation}
In Fig. 4d one can see
the $p/\pi^+$ ratio given by the fit to the data
\cite{ALEPH} (shaded area) and
the prediction of quark combinatorics (\ref{5-hep}): the agreement at
$x>0.2$ is quite good.

Let us comment the result of our calculation $p/\pi^+
\simeq 0.20$ for leading particles in jets. In the jet created by a
quark the leading hadrons are produced in proportions as it is
given by eq. (\ref{2-hep}):
$B_i:2M_i:M$.
We consider only the production of hadrons belonging to the lowest
(baryon and meson) multiplets, and, hence, keep
in (\ref{3-hep}) only the terms
with $L=0$ (hadrons from the quark $S$-wave multiplets).
In our estimations we assume
$\beta_0\simeq\mu_0$, and therefore we
substitute $B_i\to B_i(0)$, $M_i\to M_i(0)$ and $M\to M(0)$.
The precise content of $B_i(0)$, $M_i(0)$ and $M(0)$ depends on
the proportions in which the sea quarks are produced.
We assume flavour symmetry breaking for sea quarks,
$u\bar u\colon d\bar d\colon s\bar s=1 \colon 1\colon \lambda$,
with  $0\le \lambda \le 1$.
For the sake of simplicity, we put first $\lambda=0$ (actually the ratio
$p/\pi$ depends weakly on $\lambda$). Then for the $u$-quark  initiated
jet we have:
\begin{eqnarray}
&& B_u(0)\to \frac2{15}p +\frac1{15}n\ +\
(\Delta-\mbox{resonances }), \nonumber\\
&& M_u(0)\to \frac18\pi^+ +\frac1{16}\pi^0
+\frac1{16}(\eta+\eta') +(\mbox{ vector mesons }),  \nonumber\\
&& M(0)\to \frac1{16}\pi^+
+\frac1{16}\pi^0+\frac1{16}\pi^-+\frac1{16} (\eta+\eta')+(\mbox{ vector
mesons }).
\label{7-hep}
\end{eqnarray}
The hadron content of the $d$-quark  initiated jet is determined by
isotopic conjugation $p\to n$, $n\to p$, $\pi^+ \to \pi^-$, and  the
content of antiquark jets is governed by charge conjugation; in jets of
strange quarks only sea mesons ($M$) contribute to the
$p/\pi^+$ ratio.

Taking into account
the ratio $B_i:2M_i:M=1:2:1$
and the probabilities for the production of quarks of
different flavours $q_i$, given by
(1), we obtain $p/\pi^+ \simeq 0.21$ for $\lambda=0$.
We can easily get the $p/\pi^+$ ratio for an arbitrary
$\lambda$: the decomposition of the ensembles $B_i(0)$, $M_i(0)$,
$M(0)$  with respect to hadron states has been performed in Ref.
\cite{1*} (see Appendix D, Tables D.1 and D.2). But, as was stressed
above, this ratio is a weakly dependent function of $\lambda$:
at $\lambda=1$ we have $p/\pi^+ \simeq 0.20$.

\subsection{Production
of highly excited resonances and the Watson-Migdal factor }

For quark combinatorics the saturation of $q\bar q$ and $qqq$ states by
real hadrons is of principal importance. The probability of saturation
is defined by the coefficients $(\mu_L,\mu_L^{(i)})$ and
$(\beta_L,\beta_L^{(i)})$ in (72). The main question is what
contribution from high multiplets is important enough
not to be negligible in the spectra.

Consider in more detail the production of mesons in the central region:
$q\bar q \to M$. The central production of $q\bar q$ states is provided
by the diagrams of Fig. 6
(loop diagram,  Fig. 6a, and interactions of the produced quarks,
Fig. 6b).
The diagrams of the type of
Fig. 6b for final state interactions lead to the relativistic
Watson--Migdal factor. To estimate how
many highly excited states are
produced, we have to find out which states are determined by the $q\bar
q$ system in the multiperipheral ladder.

The constructive element of the ladder is a process
shown in Fig. 5b. In this process, as was stressed in Section 4, new
$q\bar q$ pairs are created at relatively large separations (in
hadronic scale), at $r\sim 1$ fm: these separations are just those
in the $q\bar q\to M$ transitions.
The orbital momenta  of the $q\bar q$ system for this
transition can be
written
as $L\sim kr$. For relative quark momenta
$k\la 0.6$ GeV/c, we have:
\be
L\la 3.
\label{75Mig}
\ee
Relying on the behaviour of the Regge trajectory, one can understand,
to what meson masses $\mu$ this relation corresponds. The trajectories
for $q\bar q$ states are
linear up to $\mu\sim 2.5$ GeV \cite{syst}:
\be
\alpha(\mu^2) \simeq \alpha(0) +\alpha'(0) \mu^2\ ,
\label{76Mig}
\ee
the slope $\alpha'(0) $ is approximately equal $\alpha'(0) \simeq
0.8$ GeV$^{-2}$ and the intercept belongs to the interval $0.25 \la
\alpha(0) \la 0.5$. Hence, for large $\mu$, the estimation gives
 $\mu^2 \sim
\alpha(\mu^2)/ \alpha'(0) $; with $\alpha(M^2)\sim 3$, we have
$\mu^2\sim 4$ GeV$^2$. So we conclude
that in the multiperipheral ladder it
would be natural to expect the production of $q\bar q$ mesons with
masses
\be
\mu\la 2 \quad {\rm GeV}.
\label{77Mig}
\ee
It is instructive to write down the $q\bar q$ states which
belong to this interval and saturate the expansions (\ref{3-hep}).
The $q\bar q$ states are characterized by the
orbital momentum $L$, the quark spin $S=0,1$ and the total momentum $J$;
one more characteristics of the $q\bar q$ state is the radial quantum
number $n$. In the region $\mu \la 2$ GeV the following $q\bar q$ multiplets
$n^{2S+1}L_J$ with $n=1$ are located \cite{PDG,syst}:
\begin{eqnarray}
&& 1^1 S_0 \ ,\quad 1^3 S_1\ ;\\ \nonumber
&& 1^1 P_1 \ ,\quad 1^3 P_J \quad  (J=0,1,2)\ ; \\ \nonumber
&& 1^1 D_2 \ ,\quad 1^3 D_J \quad  (J=1,2,3)\ ; \\ \nonumber
&& 1^1 F_3 \ ,\quad 1^3 F_J \quad  (J=2,3,4)\ .
\label{78Mig}
\end{eqnarray}
As (\ref{75Mig}) tells us, all these states
contribute significantly to the meson production.

The contribution of states with $n>1$ is determined by the structure of
the interaction of quarks in the final state, see Fig. 5b. For each
partial wave the sum of diagrams with final state interaction is:
\be
\frac1{1-B(s_{q\bar q})} \Pi(s_{q\bar q},s'_{q\bar q})
\frac1{1-B^*(s'_{q\bar q})}\ ,
\label{79Mig}
\ee
where $1/(1-B(s_{q\bar q}))$ is the relativistic Watson--Migdal factor:
\be
B(s_{q\bar q})=\int
\limits_{4m^2}^\infty \frac {ds}{\pi}
\frac {N(s)\rho(s)}{s-s_{q\bar q}}=
{\rm Re}\; B(s_{q\bar q})+{\rm i}N(s_{q\bar q})\rho(s_{q\bar q})\ .
\label{80Mig}
\ee
Here the function $N(s)$ characterizes the interaction of quarks for
given partial wave.

The cross section of $q\bar q$ pair production with invariant mass
squared $s_{q\bar q}$ is determined by (\ref{78Mig})
at $s'_{q\bar q}=s_{q\bar q}$:
\be
d\sigma \bigg (q\bar q (s_{q\bar q})+X\bigg ) \sim \Pi (s_{q\bar
q},s_{q\bar q})\,
\frac1{\left (1-B(s_{q\bar q})\right )\left (1-B^*(s_{q\bar q})
\right )} \ .
\label{81Mig}
\ee
It follows from quark--hadron duality
that $d\sigma \bigg (q\bar q (s_{q\bar q})+X\bigg ) $ describes
in the average the spectrum of produced resonance:
\be
d\sigma \bigg (q\bar q (s_{q\bar q})+X\bigg )
\simeq d\sigma\bigg ( \sum (Resonances\, near\, s_{q\bar q})+X\bigg )
.
\label{82Mig}
\ee
The averaging of the resonance production cross section is performed
over a certain vicinity of $s_{q\bar q}$. In this way we see that the
structure of the Watson--Migdal factor determines the rate of decrease of
$d\sigma \bigg (q\bar q (s_{q\bar q})+X\bigg )$ with the
growth of the invariant mass as well as the
quantitative contribution of the resonances with $n>1$.

There are the following $q\bar q$
multiplets of the radial  excitations in the region
$\mu \la 2$ GeV \cite{syst}:
\begin{eqnarray}
&& n^1 S_0 \ ,\quad n^3 S_1 \quad  (n=2,3)\ ;  \\ \nonumber
&& n^1 P_1 \ ,\quad n^3 P_{0,1,2} \quad  (n=2)\ .
\label{83Mig}
\end{eqnarray}
The probability to produce these states is determined by the
decrease of $d\sigma \bigg (q\bar q (s_{q\bar q})+X\bigg )$,
and by the Watson--Migdal factor in particular.

Summarizing, we definitely expect a considerable production of
resonances belonging to $q\bar q$ multiplets with $n=1$, see
(78). With less defiteness we can judge upon the production
of resonances with $n>1$ presented in (83): if the
Watson--Migdal factor suppresses strongly the contribution of large
$s_{q\bar q}$, the production of multiplets is suppressed as well; but
another situation is also plausible.

Only a small part of resonances belonging to the multiplets
(78) and located at $1500-2000$ MeV is included into the
tables of \cite{PDG}; a considerable number of resonances were
discovered and identified only recently, see \cite{syst,Baker,WA,AVA}
and references therein.  Moreover, numerous decay channels related to
multiparticle states have not been identified yet. This makes
the restoration procedure of the prompt  probabilities
on the basis of exclusion the known resonances \cite{7,CH,Uvar} doubtful.

There is one more effect which makes the realization of the
program of \cite{7,CH,Uvar} questionable presently: the effect of the
accumulation
of widths of the overlapping resonances
by one of them
\cite{ABS}.
An identification of broad states with widths of the order of $\Gamma
/2 \sim 400-600$ MeV in the mass region $1500-2000$ MeV looks
rather ambiguous at the present level of the experimental data,
although
the existence of such a type of states seems to be probable if we
expect the existence of exotic mesons (glueballs and hybrids) in this
mass region (see \cite{ABS} for details).

\section{Coherent production of the $\omega$ and $\rho^0$ mesons at
large $x$}

Indications on coherent phenomena in the hadronic $Z^0$ decays
follow from the ratio $\omega/\rho^0$ at $x\sim 1$, see Fig. 4f. In
multiparticle processes, where the yields of $\rho$ and $\omega$
occur without interference of flavour components, the ratio
\be
\omega/\rho^0=1
\label{6.1}
\ee
is valid.

Because of the approximate equality of masses of $\rho$ and $\omega$,
 the equation (\ref{6.1}) is not violated by the decay of the highly
excited resonances. The figure 4f demonstrates the validity of
(\ref{6.1}) at $x\sim 0.1-0.2$. However, at  $x\sim 0.5-1.0$ the ratio
$\omega/\rho^0$ is definitely less than unity, thus unambigously
pointing to the interference of flavour components of the amplitudes
$Z^0\to u\bar u$ and $Z^0\to d\bar d$.

Consider the production of $\omega$ and $\rho$ in the quark jet. If the
hadronization of quarks in jets $Z^0\to u+X$ and $Z^0\to d+X$
happens in a non-coherent way, the leading quark in the $u$-jet
picks up an antiquark $\bar u$ from the sea,  giving rise to the
transition $u\bar u \to (u\bar u+d\bar d)/2+(u\bar u-d\bar
d)/2=\omega/\sqrt{2}+\rho^0/\sqrt{2}$.
Likewise, in the $d$-jet
 $d\bar d \to (u\bar u+d\bar d)/2-(u\bar u-d\bar
d)/2=\omega/\sqrt{2}-\rho^0/\sqrt{2}$. If there is no interference of
$u$- and $d$- jets, the equation (\ref{6.1})
is valid for each of them. But if
the production amplitudes of $\omega$ and $\rho$ do interfere, the
eqaution (\ref{6.1}) is violated.  Let us consider the ratio
$\omega/\rho^0$ for such a case.

The amplitude of fragmentation production of $u$ and $d$ quarks is
determined by the vertex of the transition $Z^0\to q_i \bar q_i$ which
has the following structure  $\left [ \bar \psi_i
\gamma^\mu(g^i_V-g_A^i\gamma_5) \psi_i \right ] Z_\mu$;  for the $u$
quark $g^{(u)}_A=1/2$,  $g^{(u)}_V=1/2-4/3 \sin ^2\theta_w$ and for the
$d$ quark $g^{(d)}_A=-1/2$,  $g^{(d)}_V=-1/2+2/3 \sin ^2\theta_w$.
Therefore, for the axial-vector interaction we have the
following transition:
\be
A\to 0.5\left [u+(q\bar q-sea+\bar u) \right ]
- 0.5\left [d+(q\bar q-sea+\bar d) \right ]
\ee
$$
\to  0.5\left [(u\bar u+X_{u\bar u} \right ]-
 0.5\left [(d\bar d+X_{d\bar d} \right ]\ .
$$
If $X_{u\bar u}$ and $X_{d\bar d}$  stand for the same state,
$X_{u\bar u}=X_{d\bar d}\equiv X$, then
\be
A \to \frac12(u\bar u-d\bar d)+X=\sqrt{2}\omega +X \ .
\ee
The amplitude which is due to the vector interaction is
\be
V\to 0.19\left [u+(q\bar q-sea+\bar u) \right ]
- 0.35\left [d+(q\bar q-sea+\bar d) \right ]
\ee
$$
\to  0.19\left [(u\bar u+X_{u\bar u} \right ]-
 0.35\left [(d\bar d+X_{d\bar d} \right ]\ .
$$
With $X_{u\bar u}=X_{d\bar d}$ we have:
\be
V\to \sqrt{2}(-0.16 \omega+0.54 \rho^0)+X \ .
\ee
Therefore, for the axial-vector and vector interactions the ratios are:
\be
\left (\frac{\omega}{\rho^0}\right )_{axial}=0\ , \qquad
\left (\frac{\omega}{\rho^0}\right )_{vector}=8.8\cdot 10^{-2}\ .
\ee
Hence, the coherent processes strongly suppress the production of
$\omega$ meson compared to $\rho^0$. In the experiment
$\omega/\rho^0\simeq 0.4$ at $x\sim 0.5-1$, that enables us to evaluate
the contribution of coherent processes to be of the order of 50\%.

\section{Conclusion}

The rules of quark combinatorics are the consequence of the quark
structure of hadrons, which directly reveals itself in the ratio
$\rho_{prompt}/\pi_{prompt}=3$. This equality is valid for the
multiparticle production processes such as the hadronic $Z^0$ decays and
high--energy hadron collisions.

The spectra observed in the multiparticle production processes are
formed mainly due to the decay of highly excited resonances. In the
meson sector states with masses up to $1500-2000$ MeV contribute
significantly; we can suppose that in the baryon sector it would be
the contribution of states up to $2000-3000$ MeV. The reconstruction
 of the
spectra of highly excited resonances does not seem possible on the
present level of knowledge. Therefore, one should investigate the
prompt particle yields using jets with large $x$ where the contribution
of the resonance decay products is considerably suppressed.

The ratio $\rho^0/\pi^0=3$ observed in the hadronic $Z^0$ decay at
$x\sim 0.5-0.8$ agrees with the predictions of quark combinatorics
reasonably well.

The value  $p/\pi^+$ is also in agreement with the quark combinatorial
predictions at large $x$, thus revealing the small (about 25\%)
probability to meet the diquark loop in the chain of quark loops.

The peripheral production of new $q\bar q$ pairs in the
multiparticle production processes allows us to introduce
 the suppression parameter  for the production of strange
and heavy flavour quarks. The suppression parameters are determined by
the ratio of quark masses, their values agree with the experimental data.

In the central region, at $x \la 0.2$, the experiment gives us
$\omega/\rho^0\simeq 1$ that also agrees with the quark combinatorial
rules.  However, at $x>0.5$ the experimental data definitely tell us
that $\omega/\rho^0 < 1$ which is an indication to the interference of
flavour
amplitudes with the fragmentation production of $u$ and $d$ quarks:
the signs of flavour amplitudes favour $\omega/\rho^0 < 1$.

As a conclusion, we can state that the predictions of quark combinatorics
for hadronic
decays of the $Z^0$ boson  agree reasonably well with the experimental
data.

\section*{Acknowledgement}
The authors are deaply indebted to L.G. Dakhno for many useful
discussions.  The paper was partly supported by RFBR grant 98-02-17236.

\section*{Appendix A:
$V/P$ ratio in the high--energy hadron--hadron
collisions}

We demonstrate here that the ratio $V_{prompt}/P_{prompt}=3$ is not
specific for the decay processes ---  it is inherent in the hadronic
multiple production processes as well.

We consider the meson central production cross section, which
is governed by the two--pomeron diagram as is shown
in Fig. 7a. The probability of the meson production is determined by the
structure of a pomeron ladder; here we use a QCD-motivated pomeron
based on the pQCD pomeron of \cite{14}. In \cite{14} the pomeron was
constructed inserting the running coupling constant
and soft interaction boundary condition. The pomeron found under
these constraints,  Lipatov's pomeron \cite{14}, has an infinite set
of poles in the $j$-plane; it  is a suitable object for
using it as a guide in the analysis of the soft interaction region.
Such an extension of Lipatov's pomeron into the soft region has been
already used in \cite{csc}; below we follow the results of \cite{csc}.

The gluon ladder which responds to Lipatov's pomeron is shown in Fig.
7b. For the description of soft interactions the quark loops can be
incorporated into the gluon ladder since,
according to the rules of $1/N$ expansion
\cite{15}, the quark loops do not cause additional smallness. The
cutting of quark loops (see Fig. 7c) provides the diagram which
corresponds to
the inclusive production cross section of the $q\bar q$ pair i.e.
the production of meson states. In this sense the diagram of Fig.
7c is re-drawn as Fig. 7d: the quark--pomeron vertex is defined by
the interaction of two reggeized gluons with a quark. The spin structure
of such a vertex can be easily calculated (see \cite{csc}, Appendix C)
using the analysis of the leading-$s$ terms in reggeon
kinematics (large $s$ and small $t$) for the vertex {\it vector
particle + fermion} \cite{GLF}.

In the leading terms of the $1/N$ expansion there exist diagrams of the
type of Fig. 7e in which the gluons of the upper (or lower) pomeron
interacts with two quarks simultaneously
(generally, these diagrams can be presented
as Fig. 7f). As was shown in \cite{csc}, the sum of all these
contributions (Figs. 7d, 7e etc.) results in the colour screening
effect: the quark loop amplitude is nearly zero for the mentioned quark
configurations at $r_{q\bar q}\la 0.2$ fm (the existence  of the
colour screening effects in hadron production processes was emphasized
long ago \cite{Mue,Bro}).

The colour screening effect for the quark loop diagram in the
pomeron ladder \cite{csc} allows us to restrict the calculations
of $V_{prompt}/P_{prompt}$ to the process of Fig. 7a, because
this diagram is responsible for the main effect at
large $q\bar q$ separations, $r_{q\bar q }\ga 0.2$ fm.

The loop diagram of Fig. 7a is defined by the formula which is quite
similar to (8) --- only the spin factors are substituted as follows:
$S_\pi \to S^{(P)}_\pi$ and $S_\rho \to S^{(P)}_\rho$, that is due to
the accounting for vertices $P_{qq}$. The spin factors $S^{(P)}_\pi$
and $ S^{(P)}_\rho$ read:
\begin{eqnarray}
&& S_\pi^{(P)} =
\ -{\rm Sp}\left(i\gamma_5(\hat k_1+m)\hat n_+(\hat
k_1+m)i\gamma_5(-\hat k_2+m)(-\hat n_-)(-\hat k_2+m)\right) \nonumber\\
&& S_\rho^{(P)}
=\ -{\rm Sp}\left(\gamma^\perp_\alpha(\hat k_1+m) \hat n_+
(\hat k_1+m)
\gamma^\perp_\alpha(-\hat k_2+m)(-\hat n_-) (-\hat k_2+m) \right)\ .
\label{A.1}
\end{eqnarray}
Here $\hat n_+$ is the quark--pomeron vertex (the coupling to the upper
pomeron in Fig. 7a) and $(-\hat n_-)$ is the antiquark--pomeron vertex
(the coupling to the lower pomeron);
we have also taken into account that pomerons in the diagram of
Fig. 7a for the inclusive cross section carry zero momenta.

If the upper pomeron in the graph of Fig. 7a moves along the $z$-axis
with momentum $p_A=(p_0,0,p_z)\simeq (p_z,0,p_z)$, while the lower
one moves in opposite direction, we have
\be
n_-=(1,0,-1), \qquad n_+=(1,0,1)\ .
\label{A.2}
\ee
As in Section 2, the loop diagram has been calculated in
terms of the spectral
integration; because of that $k_1^2=k_2^2=m^2$. After the operators
$\hat n_+$ and $\hat n_-$ have been commutated with $\hat k_1$ and
$\hat k_2$, we have:
\begin{eqnarray}
&& S_\pi^{(P)} =
\ -{\rm Sp}\left(i\gamma_5(\hat k_1+m)
i\gamma_5(-\hat k_2+m)\right)
4(n_+ k_1)( n_- k_2)
\nonumber\\
&& S_\rho^{(P)}
=\ -{\rm Sp}\left(\gamma^\perp_\alpha
(\hat k_1+m)
\gamma^\perp_\alpha(-\hat k_2+m) \right)\
4( n_+ k_1)( n_- k_2)
\label{A.3}
\end{eqnarray}
Up to a common factor, $S^{(P)}_\pi$ and $S^{(P)}_\rho$ coincide with
the normalizing spin factors of the pion and $\rho$-meson wave functions
given by (24) and (29):
\begin{eqnarray}
&& S_\pi^{(P)} =
4( n_+ k_1)( n_- k_2)
 S_\pi^{(wf)}
\nonumber\\
&& S_\rho^{(P)}
=4( n_+ k_1)( n_- k_2) \; 3
S_\rho^{(wf)} \; .
\label{A.4}
\end{eqnarray}
This equation demonstrates that the two--pomeron diagram of Fig. 7a
which defines the inclusive production cross section  of $\rho $ and
$\pi$ in the central region of hadron--hadron collisions gives us
$\rho_{prompt}/\pi_{prompt}=3$.

\section*{Appendix B: Spin factors for the fragmentation production}

Below the  explicit expressions for $S^{(fr)}_P/S^{(wf)}_P$ and
$S^{(fr)}_V/S^{(wf)}_V$ are given for the processes
$Z^0 \to b\bar b\to B+X$ and
$Z^0 \to b\bar b\to B^*+X$.

For the pseudoscalar particle we have:
\begin{equation}
\frac{S^{(fr)}_P}{S^{(wf)}_P}=A_1+R^2\;A_2\ ,
\end{equation}
where
\begin{eqnarray}
A_1&=&\left ((m_b-m)^2-s'\right )(M^2+2m^2_b)
     +\left ((m_b-m)^2-s \right )(M'^2+2m^2_b)\ ,
\nonumber \\
A_2&=& M^2 \left(1-\frac{m^2_b}{M'^2}\right )\left ((m_b-m)^2-s'\right )
      +M'^2\left(1-\frac{m^2_b}{M^2} \right )\left ((m_b-m)^2-s\right )
\nonumber \\
      &-&3m^2_b\left ((m_b-m)^2-s'\right )
      -3m^2_b\left ((m_b-m)^2-s\right )\ ;
\end{eqnarray}
and for the fragmentation production of the vector particle:
\begin{eqnarray}
&&\frac{S^{(fr)}_V}{S^{(wf)}_V}
=\frac{M^2}{ss'}A_3+\frac{M'^2}{ss'}A_4+
\frac{2m^2}{ss'}(A_3+A_4)
\nonumber \\
&&+\frac{R^2}{M'^2M^2s's}\left ( M^4(M'^2-m^2_b)A_3
                          +M'^4(M^2-m^2_b)A_4
                          -3M'^2M^2m^2_b(A_3+A_4)\right ) .
\end{eqnarray}
Here
\begin{eqnarray}
A_3&=&(s+s')(m^2_b-m^2)^2
+ss'(3m^2-m^2_b-10m_bm-4s')
\nonumber \\
&+&s'^2(3m^2_b-2m_bm-m^2)  \ ,
\nonumber \\
A_4&=&(s+s')(m^2_b-m^2)^2
+ss'(3m^2-m^2_b-10m_bm-4s)
\nonumber \\
&+&s^2(3m^2_b-2m_bm-m^2) \ .
\end{eqnarray}

\begin{figure}
\centerline{\epsfig{file=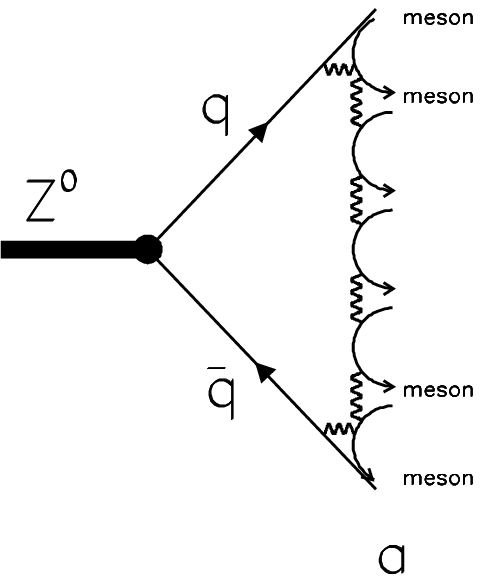,width=6.0cm}\hspace{2cm}
            \epsfig{file=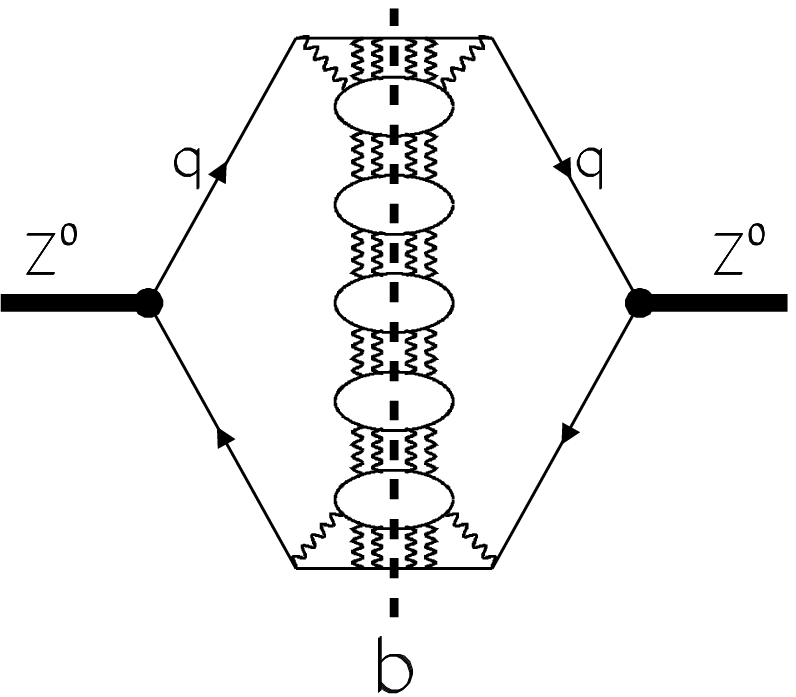,width=6.0cm}}
\vspace{1cm}
\centerline{\epsfig{file=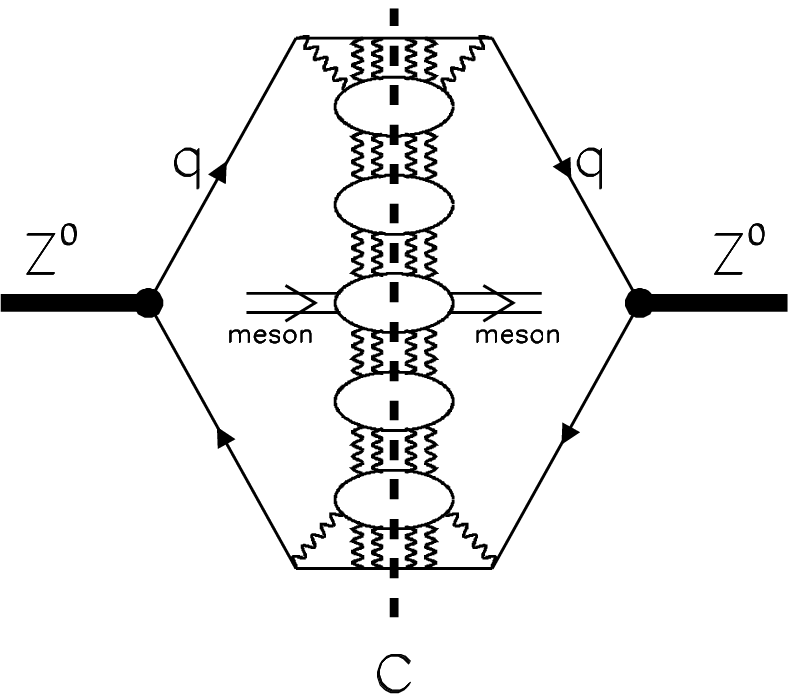,width=6.0cm}\hspace{2cm}
            \epsfig{file=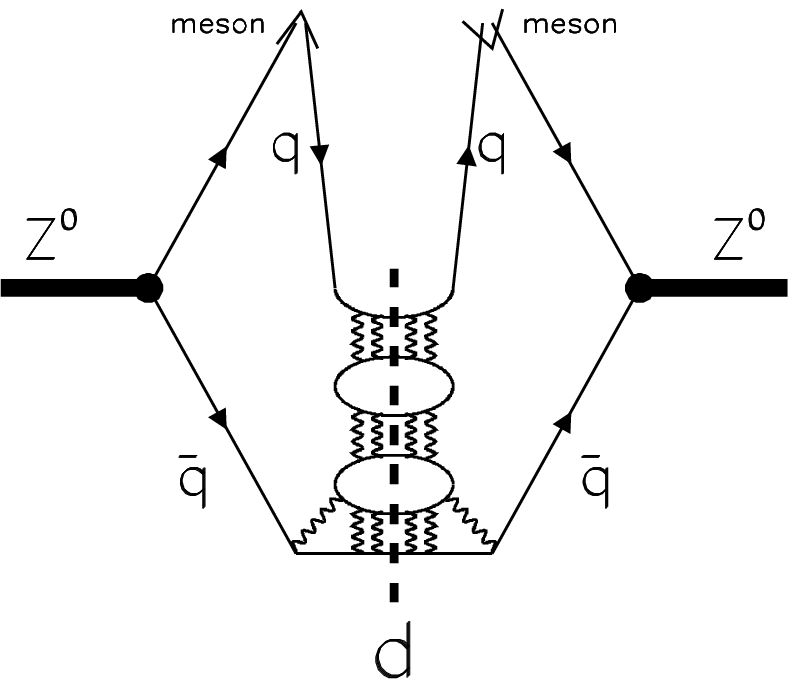,width=6.0cm}}
\vspace{1cm}
\caption{Hadronic decay of $Z^0$ meson. a) Multiperipheral ladder which
provides the transfer of colour from quark to antiqurk. b) Self--energy
diagram; the cutting along hadronic states (dashed line) determines the
hadronic cross section. c) Diagram for the inclusive meson production
cross section in the central region, $d\sigma(Z^0\to meson+X)/dx$ at
$x\sim 0$. d) Diagram for the inclusive meson production
cross section in the fragmentation region at $x\sim 0.2 -1.$}
\end{figure}

\newpage
\begin{figure}
\centerline{\epsfig{file=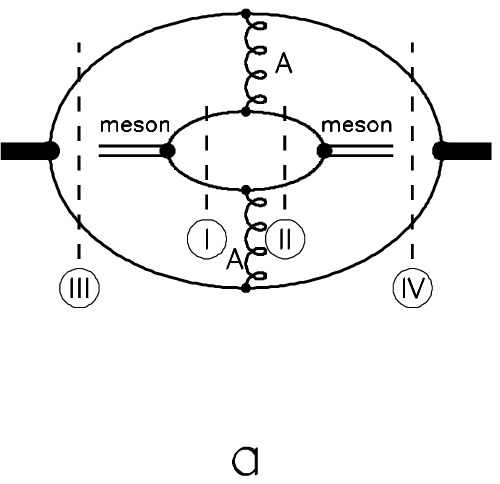,width=5.0cm}\hspace{1cm}
            \epsfig{file=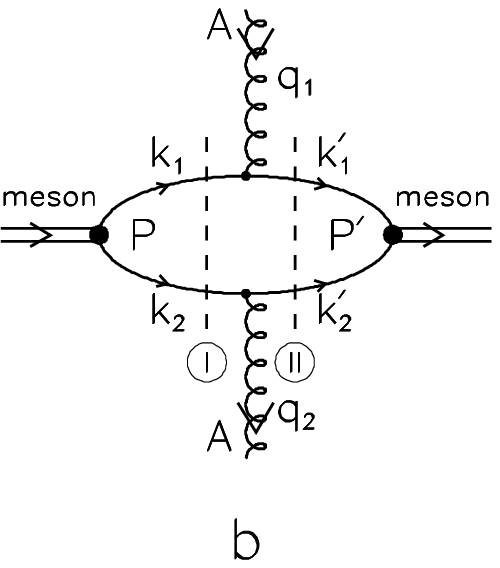,width=5.0cm}}
\centerline{\epsfig{file=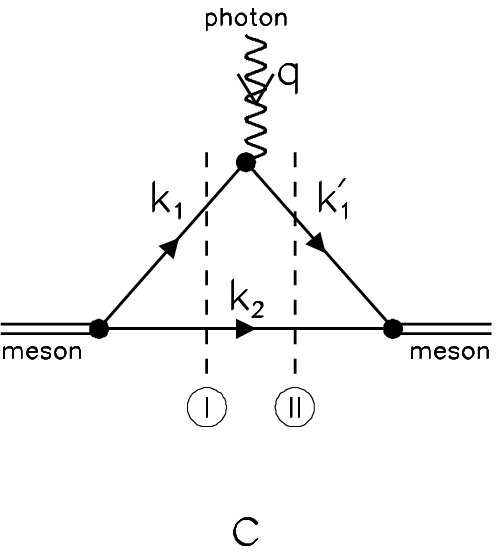,width=5.0cm}\hspace{1cm}
            \epsfig{file=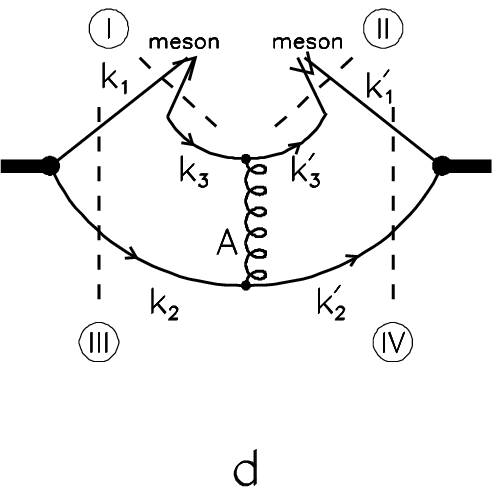,width=6.0cm}}
\vspace{1cm}
\caption{a,b) Production blocks for meson in the central region; the
dashed lines show the cuttings of diagrams in the spectral integral.
c) Form factor diagram determining the meson wave function.
d) The cuttings of the diagram for the inclusive meson production cross
section in the fragmentation region.}
\end{figure}

\newpage
\begin{figure}
\centerline{\epsfig{file=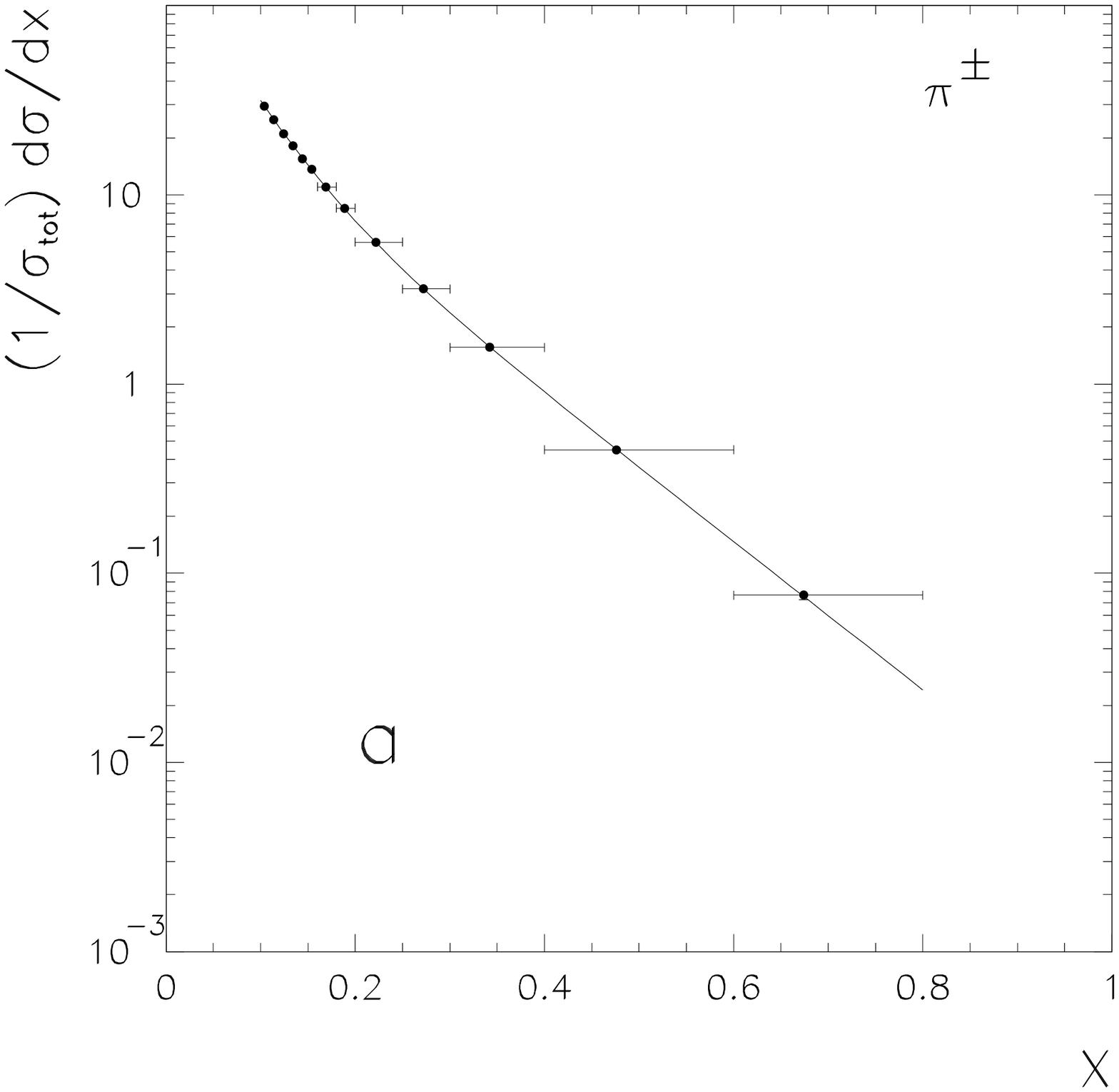,width=8.0cm}
            \epsfig{file=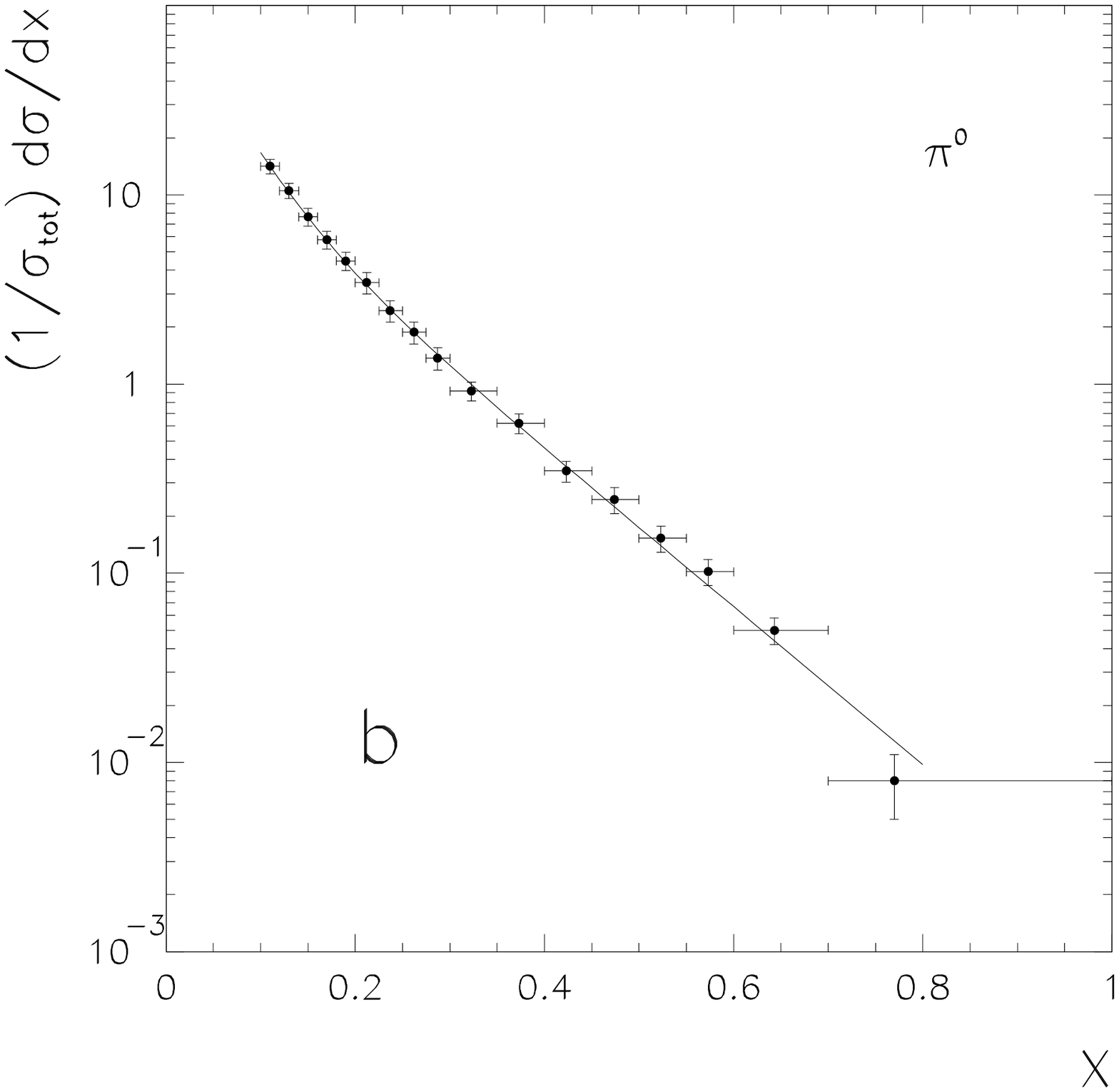,width=8.0cm}}
\centerline{\epsfig{file=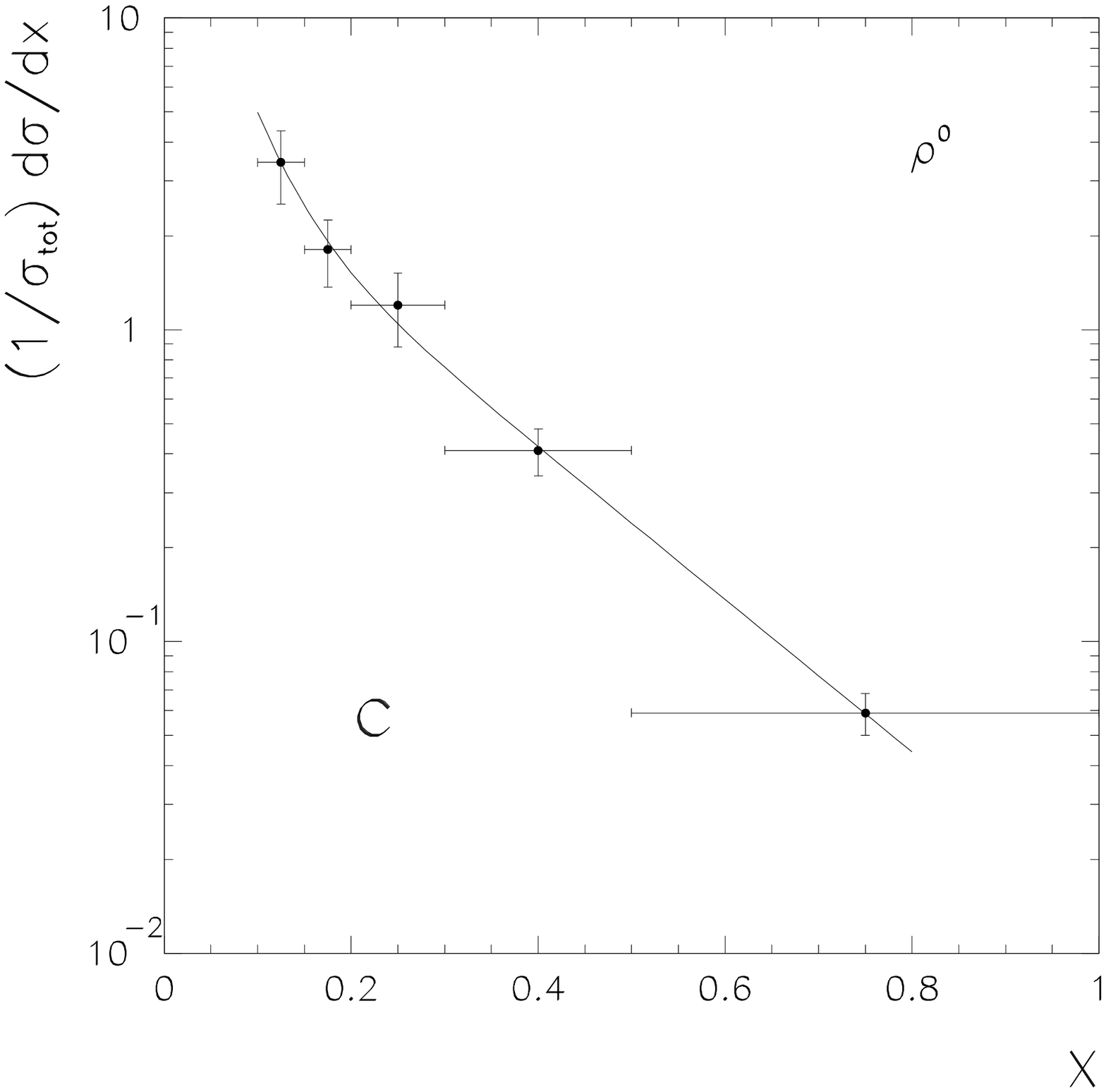,width=8cm}
            \epsfig{file=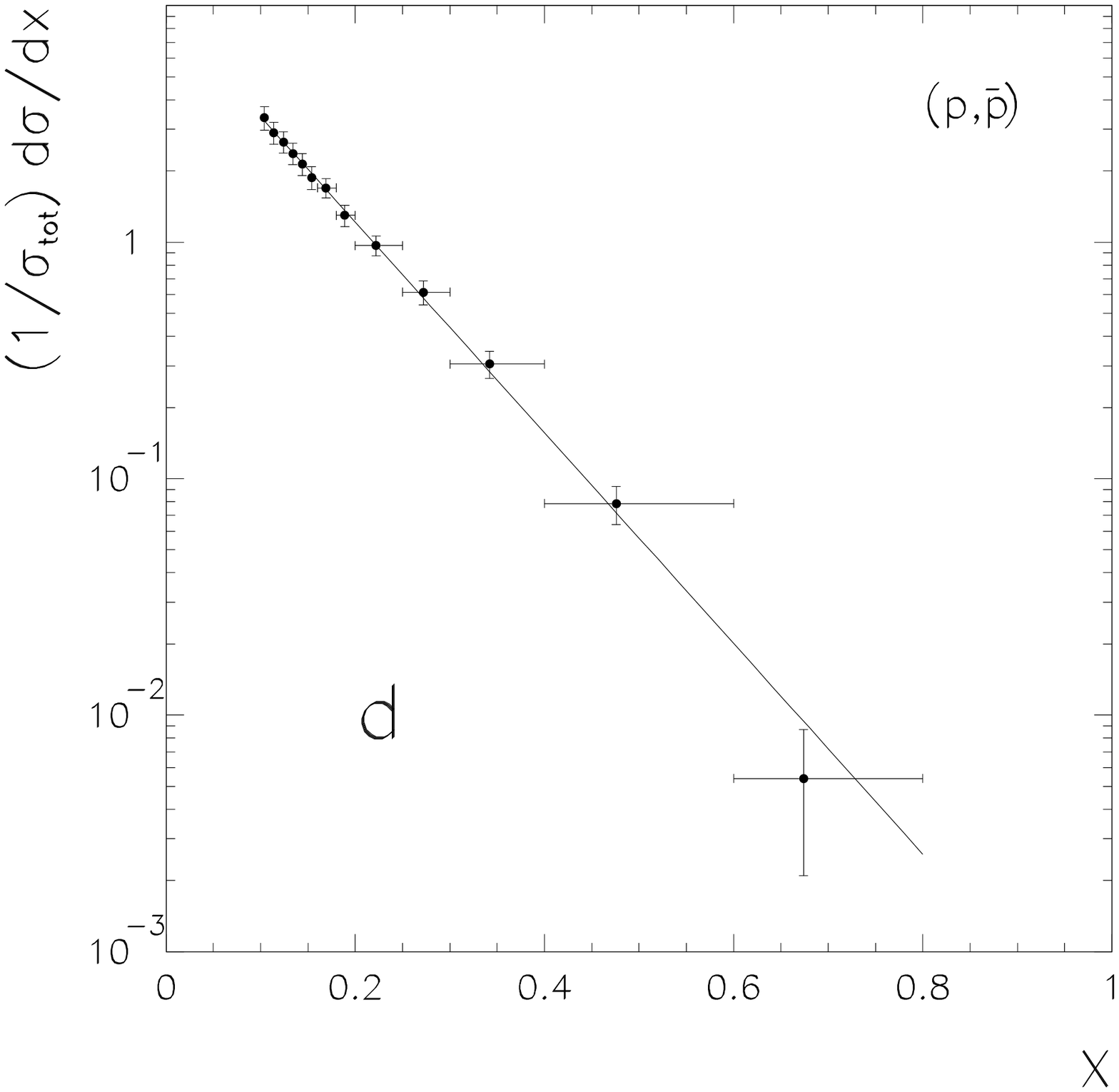,width=8cm}}
\caption{The spectra $(1/\sigma_{hadron})\; d\sigma_{Z^0\to meson+X}/dx$
[11]
for a) $\pi^\pm$, b) $\pi^0$, c) $\rho^0$ and d) $(p,\bar p)$ and their
fit by exponential functions $\Sigma C_i \exp(-b_ix)$ [11].}
\end{figure}

\newpage
\begin{figure}
\centerline{\epsfig{file=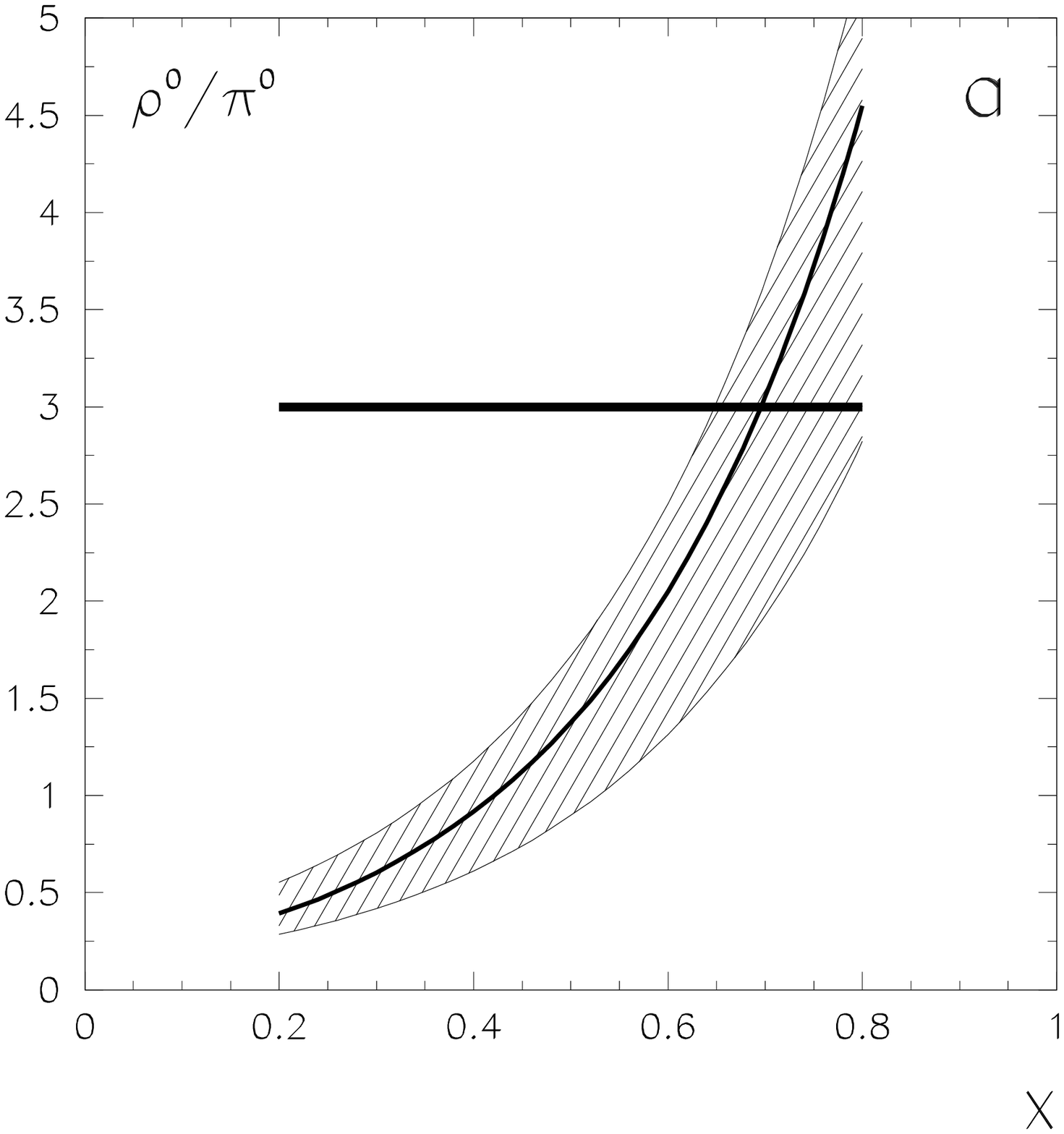,width=6.5cm}
            \epsfig{file=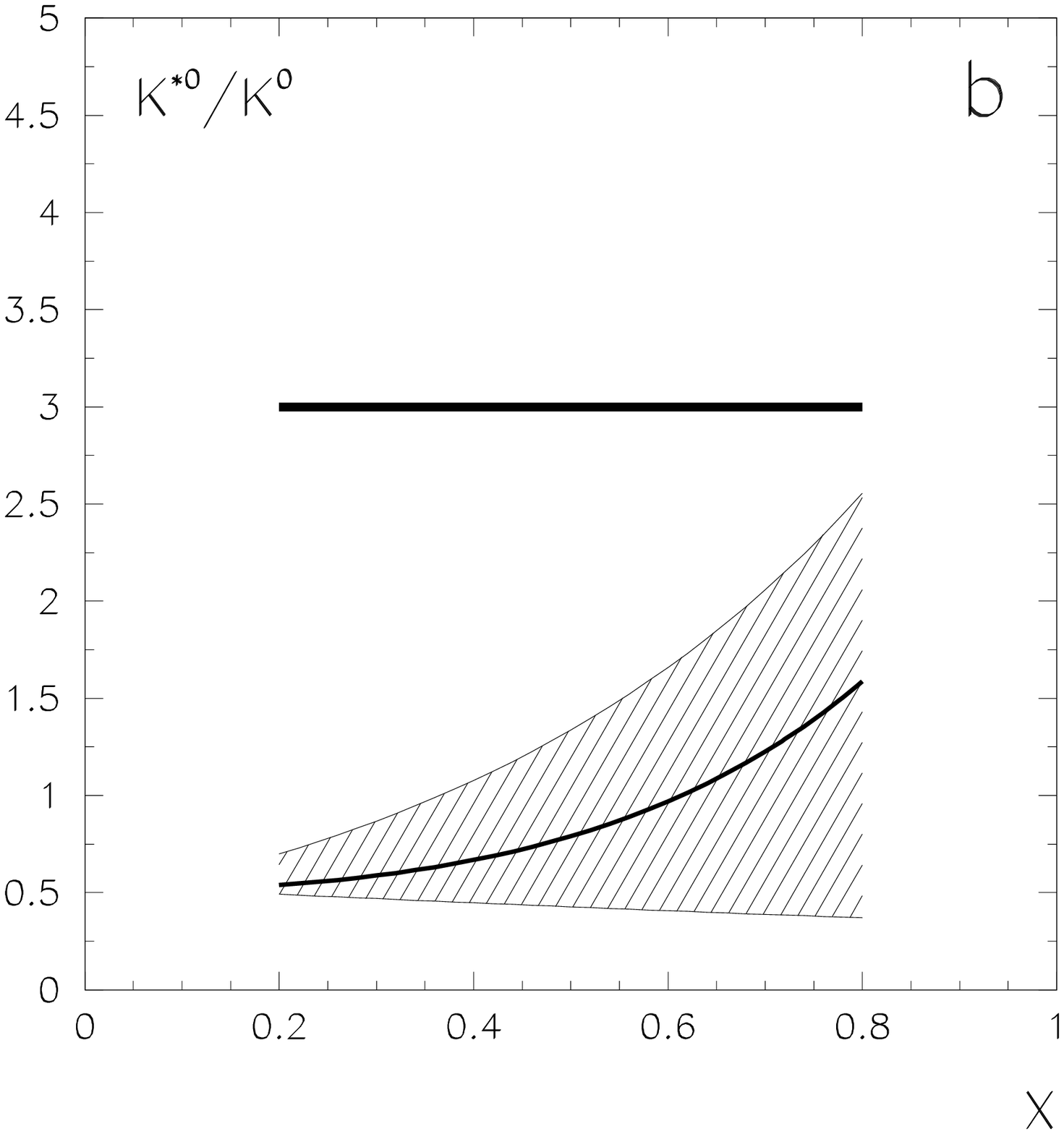,width=6.5cm}}
\vspace{-0.5cm}
\centerline{\epsfig{file=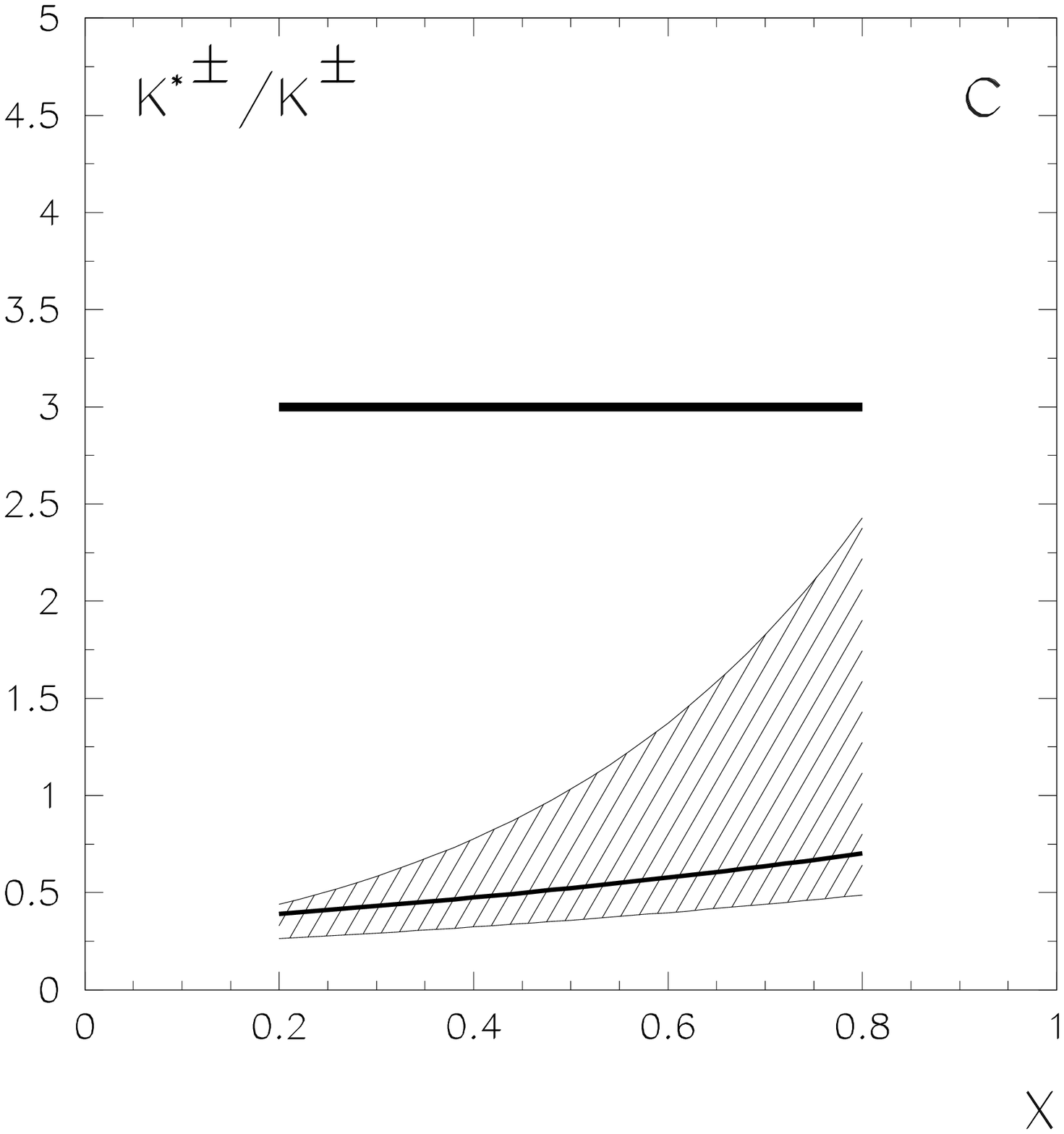,width=6.5cm}
            \epsfig{file=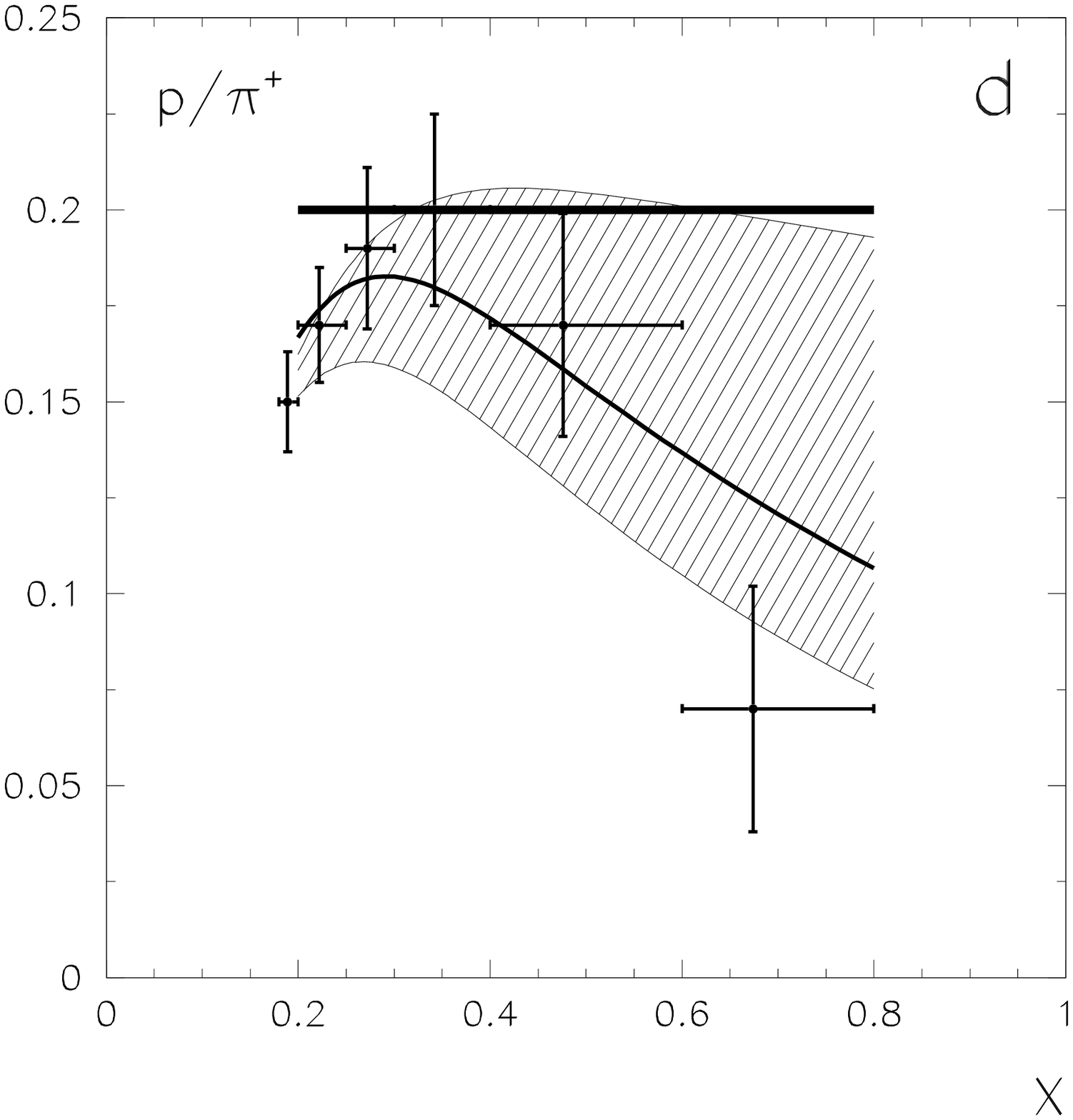,width=6.5cm}}
\vspace{-0.5cm}
\centerline{\epsfig{file=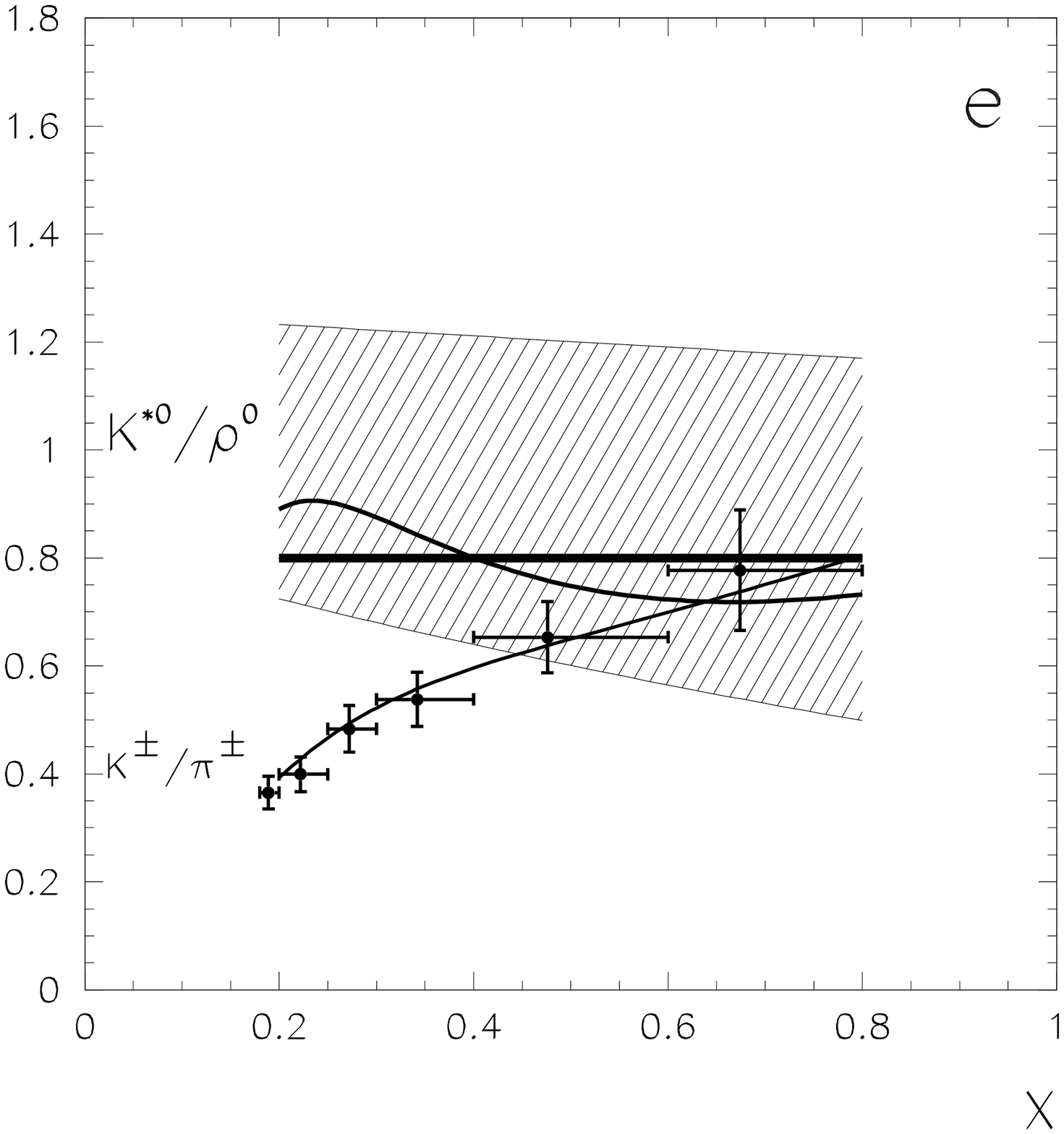,width=6.5cm}
            \epsfig{file=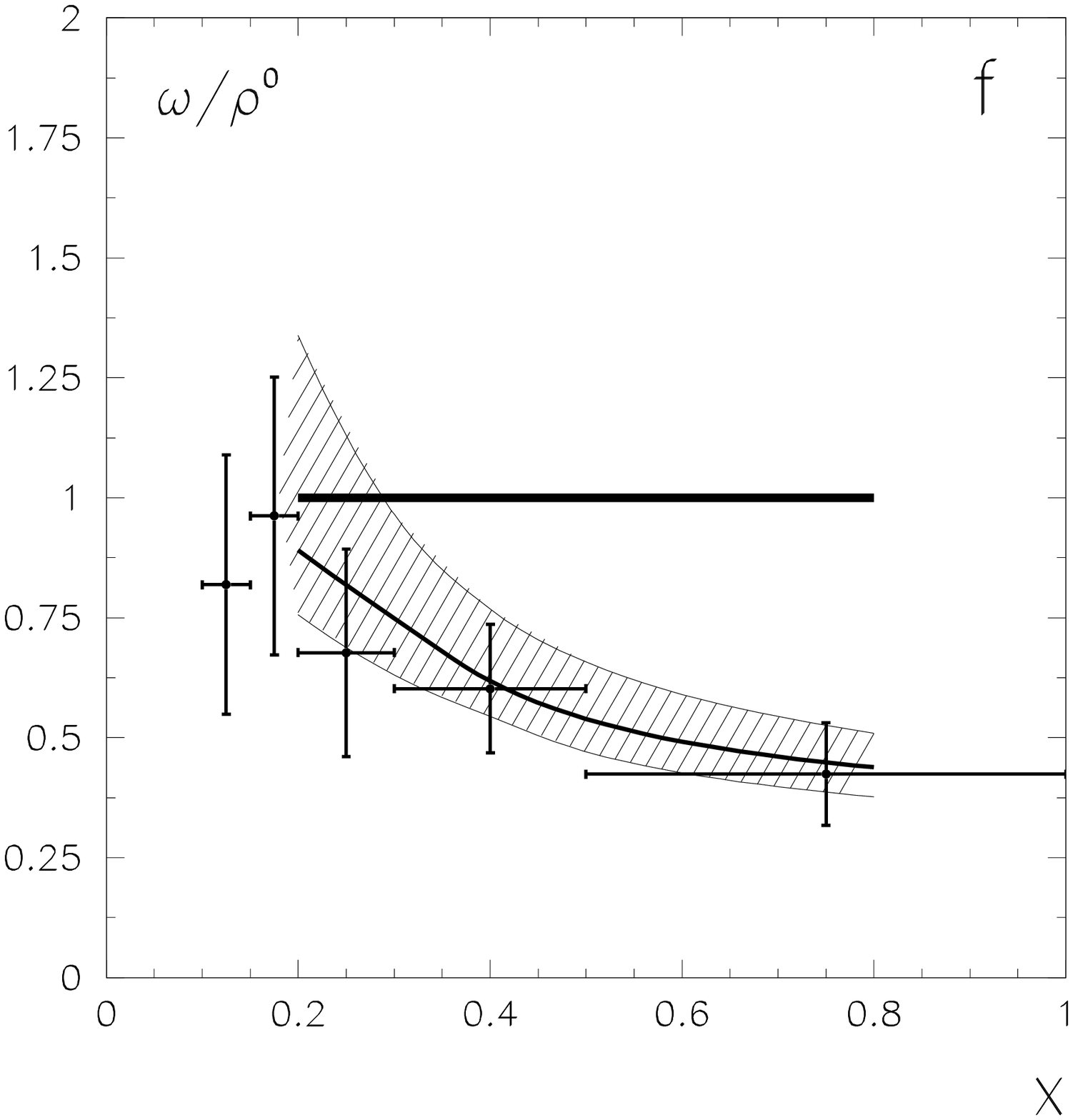,width=6.5cm}}
\caption{Ratios $\rho^0/\pi^0$ (a), $K^{*0}/K^0$ (b),
$K^{*\pm}/K^\pm$ (c), $p/\pi^+$ (d), $K^{*0}/\rho^0$ and
$K^{\pm}/\pi^\pm$ (e), $\omega/\rho^0$ (f) obtained from the
exponential fit to ALEPH data [11] (shaded areas). Solid lines stand for
the predictions of quark combinatorics rules. In Figs. 4d, 4e, 4f the
ratios $p/\pi^+$, $K^{\pm}/\pi^\pm$ and $\omega/\rho^0$
which are obtained using a histogram description of the
spectra [11] are also shown.}
\end{figure}

\newpage
\begin{figure}
\centerline{\epsfig{file=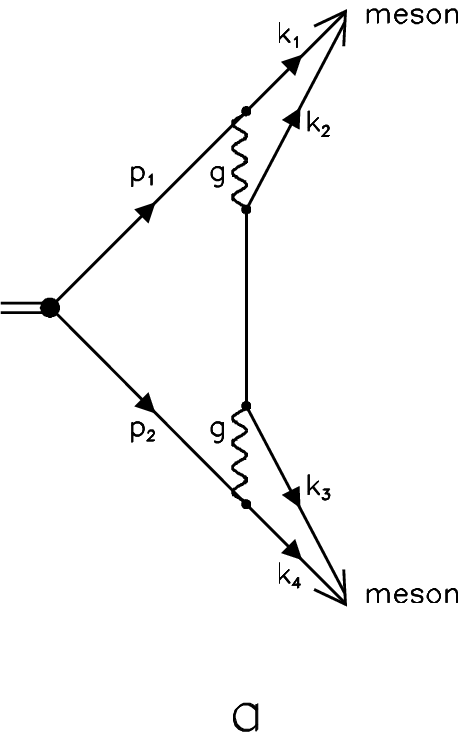,width=6.0cm}\hspace{1cm}
            \epsfig{file=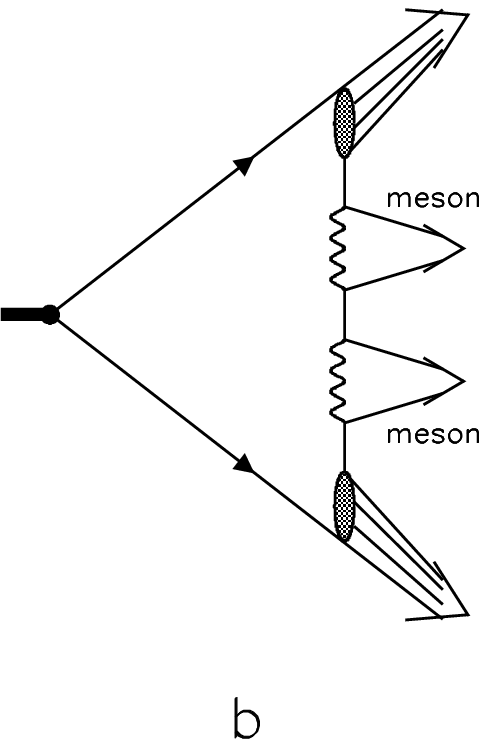,width=6.0cm}}
\vspace{1cm}
\centerline{\epsfig{file=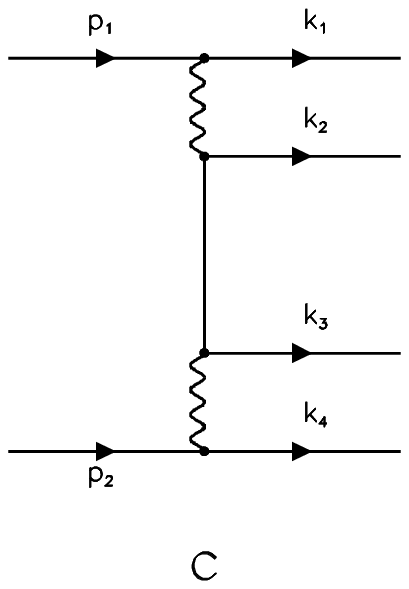,width=6.0cm}}
\caption{a) Meson decay diagram. b) The transition $q\bar q\to meson +
meson$  as the constructive element of the multiperipheral chain in the
hadronic $Z^0$ decay. c) Diagram for the peripheral production of the
new $q\bar q$ pair.}
\end{figure}

\newpage
\begin{figure}
\centerline{\epsfig{file=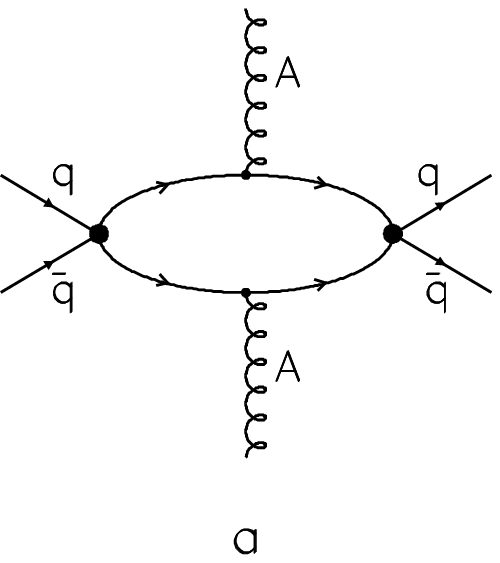,height=5.0cm}\hspace{1cm}
            \epsfig{file=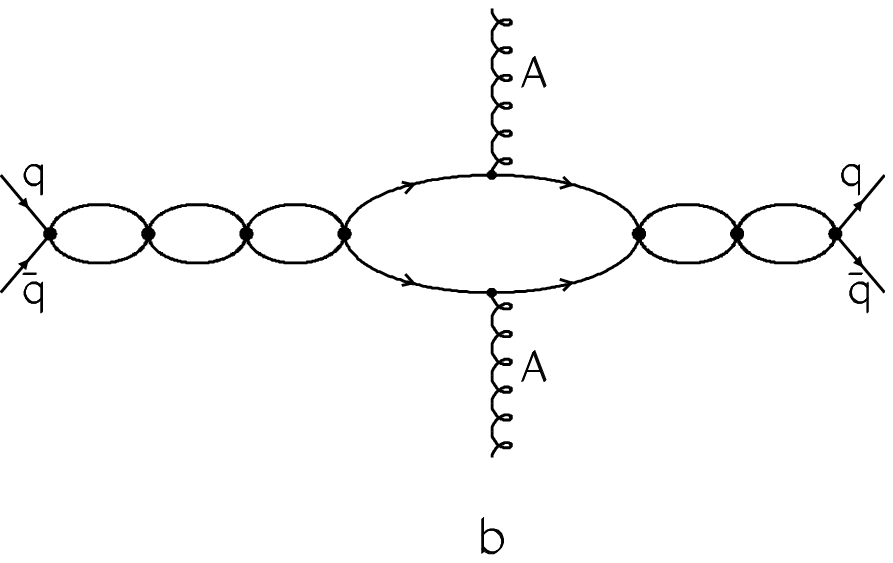,height=5.0cm}}
\vspace{1cm}
\caption{Diagrams for the central production of $q\bar q$ pair. The set
of loop diagrams responsible for the Watson--Migdal factor.}
\end{figure}

\newpage
\begin{figure}
\centerline{\epsfig{file=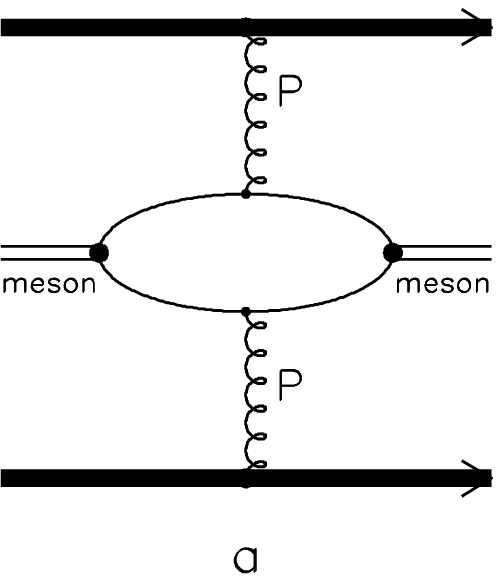,width=5.0cm}\hspace{1cm}
            \epsfig{file=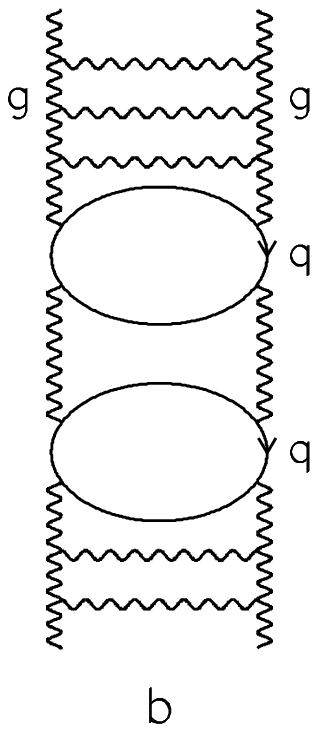,width=5.0cm}\hspace{1cm}
            \epsfig{file=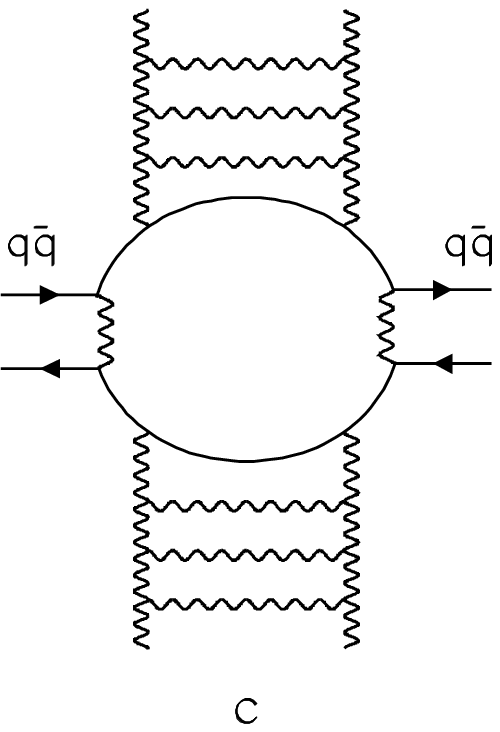,width=5.0cm}}
\vspace{1cm}
\centerline{\epsfig{file=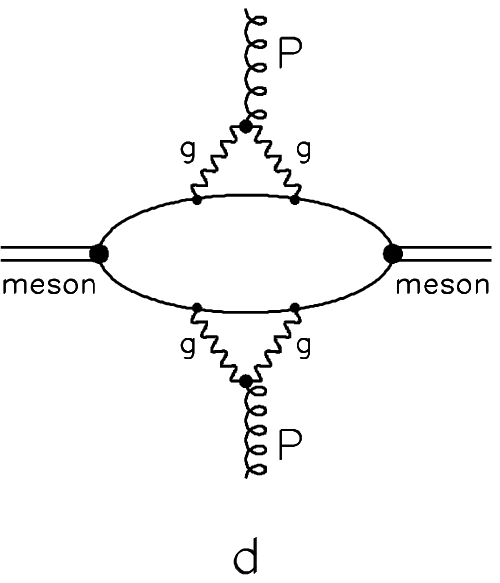,width=5.0cm}\hspace{1cm}
            \epsfig{file=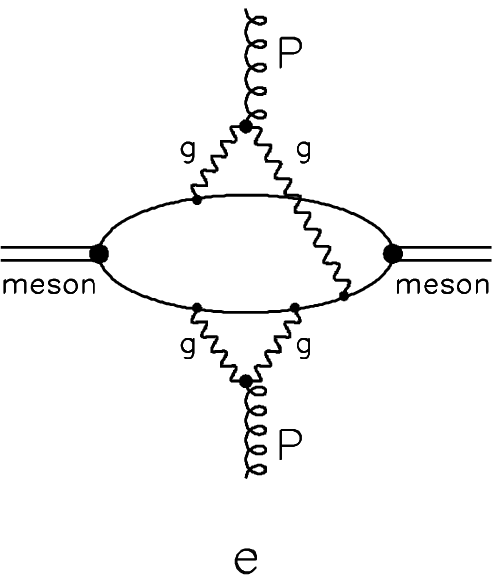,width=5.0cm}\hspace{1cm}
            \epsfig{file=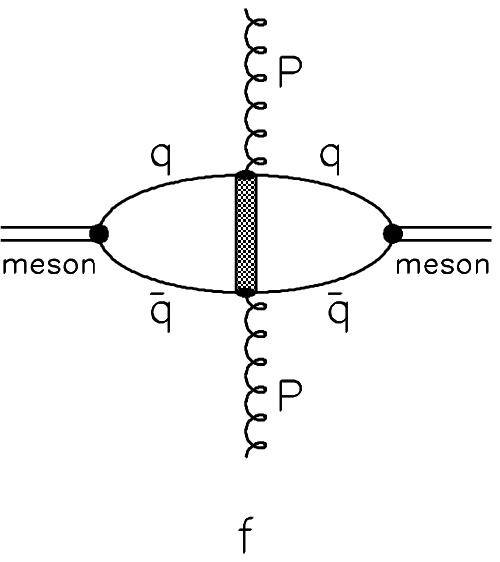,width=5.0cm}}
\vspace{1cm}
\caption{Diagrams for meson production in the central region in high
energy hadron--hadron collisions.}
\end{figure}

\end{document}